\documentclass{ametsoc}
\usepackage[]{lineno}

\usepackage{enumerate}

\nolinenumbers
\journal{jas}

\bibpunct{(}{)}{;}{a}{}{,}

\title{What determines the distribution of shallow convective mass flux \\ through cloud base?}

\authors{Mirjana Sakradzija \correspondingauthor{Max Planck Institute for Meteorology,  
Bundesstr. 53, 20146 Hamburg, Germany. \email{mirjana.sakradzija@mpimet.mpg.de} } 
and Cathy Hohenegger }

\affiliation{Max Planck Institute for Meteorology,  Bundesstr. 53, 20146 Hamburg, Germany.}

\abstract{The distribution of cloud-base mass flux is studied using large-eddy simulations (LES) of two reference cases, 
	one representing conditions over the tropical ocean, and another one representing mid-latitude conditions over land. 
	To examine what sets the difference between the two distributions, nine additional LES cases are set up as variations
	of the two reference cases. We find that the total surface heat flux and its changes over the diurnal cycle do not influence 
	the distribution shape. The latter is also not determined by the level of organization in the cloud field. It is instead 
	determined by the ratio of the surface sensible heat flux to the latent heat flux, the Bowen ratio $B$. $B$ sets the 
	thermodynamic efficiency of the moist convective heat cycle, which determines the portion of the total surface heat 
	flux that can be transformed into mechanical work of convection against mechanical dissipation. The thermodynamic 
	moist heat cycle sets the average mass flux per cloud $\langle m \rangle$, and through $\langle m \rangle$ it also 
	controls the shape of the distribution. An expression for $\langle m \rangle$ is derived based on the moist convective 
	heat cycle and is evaluated against LES. This expression can be used in shallow cumulus parameterizations as a physical 
	constraint on the mass flux distribution. The similarity between the mass flux and the cloud area distributions indicate 
	that $B$ also has a role in shaping the cloud area distribution, which could explain its different shapes and slopes 
	observed in previous studies. }
\begin{document}

\maketitle

%

\section{Introduction}

Since the seminal work on parameterization of cumulus clouds by \cite{arakawa-schubert:1974}, AS-74, 
the understanding of the spectral distribution of cloud properties and how it is controlled by the large-scale 
environment remains an obstacle for the formulation of convection parameterizations. In their paper 
AS-74 wrote: "Our final problem is to find the mass flux distribution function. The real conceptual difficulty 
in parameterizing cumulus convection starts from this point. We must determine how the large-scale processes 
control the spectral distribution of clouds, in terms of the mass flux distribution function, if they indeed do so. 
This is the essence of the parameterization problem."  With this in mind, it is the goal of our paper to determine 
how the mass flux distribution of shallow cumulus clouds $p(m)$ is controlled by the underlying physical processes 
and large-scale conditions. 

In the formulation of the AS-74 parameterization, the mass flux distribution function refers to the spectral distribution 
of cloud subensembles. The subensembles encompass clouds of different types based on their sizes and cloud top 
heights. This distribution is estimated in AS-74 by numerical solution of the Fredholm integral equation assuming 
convective quasi-equilibrium (QE). Here, we instead regard the mass flux distribution as an asymptotic distribution 
of the spectral subensembles that are reduced to single clouds, which then can be classified as a cloud population 
distribution. In this way, we approach the problem from another point of view: instead of assuming convective QE 
and solving for the spectral distribution of mass fluxes numerically, we focus on the underlying physical principles 
that determine the shape of $p(m)$ and its parameters. 

The decision to examine the population distribution $p(m)$ instead of the spectral distribution based on cloud 
types comes from the need to formulate a scale-aware parameterization. As the model resolution increases to 
the kilometre scale, the separation of the cloud ensemble into spectral bins that represent clouds of different 
types loses statistical significance. Instead, a cloud sample within a grid box can be viewed as a random sample 
of clouds drawn from the cloud population. The clouds are grouped by the grid box boundaries regardless of 
the cloud types. The total mass flux in a grid box $M$ is then a sum over the sampled clouds, $M=\sum^n_{i=1} m_i$, 
and its spatial distribution $p(M)$ is characterized by a spectrum of shapes starting from a normal-like distribution 
on the coarse grids, toward a long-tailed distribution on the kilometre-scale grids  \citep{craig-cohen:2006, sakradzija:2015}. 
The distribution of the total mass flux within model boxes $p(M)$ has been parameterized based on the principles 
of statistical-mechanics, and has been applied to deep convection by \cite{pc:2008}, and further developed to  
a parameterization of shallow convection by \cite{sakradzija:2015,sakradzija:2016}. In the context of such a 
parameterization, it is important to understand the physical constraints on $p(m)$ because fluctuations of the 
subgrid-scale convective tendencies influence convective regimes, organization as well as energetics of the 
explicitly modelled atmospheric flows \citep[][]{sakradzija:2016}. 

The evidence about $p(m)$ based on observations is not extensive. A few observational studies that examined 
$p(m)$ among other cloud statistics were focused on cumulonimbus clouds, for which $p(m)$ was fitted to a 
log-normal distribution function \citep{lemone-zipser:1980, jorgensen-lemone:1989}. More evidence about 
$p(m)$ has been provided by modelling studies using cloud-resolving models (CRM) or large-eddy simulations 
(LES). In a CRM study of an equilibrium deep-convective ensemble under homogeneous large-scale forcing, $p(m)$ 
was fitted to an exponential function \citep{cohen-craig:2006}. This fit was supported by theoretical derivation 
using the formalism of the Gibbs canonical ensemble from statistical mechanics \citep{craig-cohen:2006}. As more 
computing power allowed performing simulations with resolutions on the order of 100~m, it was revealed that the 
shape of this distribution is dependent on the horizontal resolution. With kilometre-scale resolution, where the 
deep cumulus clouds are not fully resolved, $p(m)$ takes an exponential-like shape, while the shape changes 
towards a power-law distribution when using higher resolution \citep{scheufele:2014}. \cite{scheufele:2014} further 
demonstrated that the power-law-like shape emerges as a result of self-organization of the individual cloud updrafts. 

For shallow cumulus clouds over the ocean, \cite{sakradzija:2015} found that the overall shape of the mass flux 
distribution results from the superposition of two distribution modes, one corresponding to the active buoyant 
clouds and the other one to non-buoyant clouds. The two modes of the cumulus cloud distribution deviate from 
an exponential shape due to correlation between cloud mass fluxes and cloud lifetimes. Each mode can be described 
using a Weibull distribution with two parameters, shape $k$ and scale $\lambda$ \citep[see Eq.~\ref{weibull}, and also][]{sakradzija:2015}. In the 
case of shallow cumulus clouds, the shape parameter of the Weibull distribution is less than one, $k<1$, which 
signifies that it is a heavy-tailed distribution. The combination of at least two Weibull distribution modes results in 
a distribution of the shallow cumulus mass flux that takes an overall power-law-like shape (see section 3). Hence, 
it appears that different mechanisms can lead to power-law distributions \citep[see e.g.][]{mitzenmacher2003, newman:2005}. 
Moreover, either a power-law or a log-normal distribution can be generated by the same underlying mechanism 
under slightly different conditions \citep[e.g.][]{mitzenmacher2003} and it is often difficult to rule out one or the 
other functional form. 

It might be possible to gain more insight into the mass flux distribution $p(m)$ by making a parallel to the distribution 
of cloud sizes. Based on the findings of modeling and observational studies, there is no consensus on the functional 
form that best describes the cloud size distribution. The suggested functions span from exponential 
\citep{plank:1969,hozumi:1982,astin-latter:1998}, over log-normal \citep{lopez:1977, lemone-zipser:1980, jorgensen-lemone:1989} 
to power-law functions with single \citep{LOVEJOY185,zhao-digirolamo:2007,wood-field:2011,dawe-austin:2012} 
or double slopes \citep{cahalan-joseph:1989, segunpta:1990, nair:1998, benner-curry:1998,neggersetal:2003,trivej-stevens:2010,heus-seifert:2013}. 
Most studies, in particular more recent ones, suggest power-laws, with or without a break in the power-law scaling at the intermediate 
cloud sizes. This scale break manifests itself as a change in the slope of a power-law distribution or as an exponential cut-off near the distribution tail. 
However, no explanation supported by evidence has been provided for the observed differences in the 
distribution shapes and slopes, and some of these differences may just reflect different meteorological conditions. 

Given that the characteristics of cloud updrafts are substantially different between tropical oceanic and midlatitude 
continental cumulus convection \citep{xu-randal:2001}, the dependency of $p(m)$ on meteorological conditions 
is not surprising. We nevertheless suspect that there are some dominant macroscopic parameters or processes 
that determine the characteristic cloud size and the mass flux that cause the variations in $p(m)$ between different cases 
and locations. Instead of assuming a distribution functional form and estimating the distribution parameters by statistical 
fitting of modeled or observed clouds, we set out to identify the physical mechanisms that might lead to a specific 
distribution functional form and a characteristic scale. We use LES of shallow cumulus convection based on two 
measurement campaigns, RICO (Rain In Cumulus over the Ocean) to represent conditions over the ocean, and 
measurements in an ARM (Atmospheric Radiative Measurements) site to represent conditions over land (Section 2). 
We aim to reveal what makes the difference in $p(m)$ between these two reference cases and to derive a 
parameterization for the distribution parameters that applies to oceanic and land conditions. 

In nine additional simulations, the two reference cases are modified (see section 2) to test the impacts of the large-scale 
forcing and surface conditions on $p(m)$. Cloud lifecycles are studied using the method of cloud tracking, also described 
in Section 2. This method provides the lifetime-averaged cloud mass flux distribution defined in section 3. Several reasons for 
the difference in $p(m)$ between the two reference cases are hypothesized and tested in section 4. In section 5 we describe 
the physical principle that explains the difference between the two characteristic distribution shapes.  
The distribution is then fitted to the mixed Weibull function to estimate the remaining unknown parameters (section 6). 
Conclusions are given in section 7.

\section{LES case studies}

Simulations were performed using the University of California, Los Angeles, large-eddy simulation (UCLA-LES) model 
\citep{stevens:1999, stevens:2010}. A detailed description of the UCLA-LES model and the specification of the parameters 
and constants used in our study are provided in \cite{stevens:2010}. The UCLA-LES model solves the Ogura-Phillips 
anelastic equations, discretised over the doubly periodic uniform Arakawa C-grid. The prognostic variables include 
the wind components~$u,v$ and $w$, liquid water potential temperature $\theta_l$, total water mixing ratio $q_t$, 
and in the precipitating cases (see the next paragraph), rain mass mixing ratio $q_r$ and rain number mixing ratio 
$N_r$. In the precipitating cases, the double-moment warm-rain scheme of \cite{seifert:2001} is used to compute 
the cloud microphysics. The subgrid turbulent fluxes are computed using the Smagorinsky-Lilly scheme 
\citep[as described in][]{stevens:1999, stevens:2010}. A third-order Runge-Kutta method is used for numerical time 
integration, a directionally split monotone upwind scheme is used for the advection of scalars, and directionally split 
fourth-order centered scheme is used for the momentum advection \citep[see][]{stevens:2010}. The effects of radiation 
are prescribed as net forcing tendencies. 

As a first reference case (R-base), an LES case study of shallow convection based on the Rain In Cumulus over the Ocean 
(RICO) measurement campaign \citep{rauber:2007} is used to represent conditions over the tropical ocean. The field 
measurements were taken during the winter season 2004/2005 in the trade-wind region of the Western Atlantic upwind 
of the islands of Antigua and Barbuda \citep{rauber:2007}. The initial profiles of potential temperature $\theta$, specific 
humidity $q_v$ and the horizontal winds $u$ and $v$ are constructed as piece-wise linear fits of the averaged profiles 
from the radiosonde measurements taken over Barbuda during a period with no disturbance due to mesoscale convective 
systems \citep[Fig.~2 and Table 2 in][]{vanzanten:2011}. Vertical time-invariant profiles of the subsidence rate and of 
horizontal advection of moisture and temperature are prescribed and act on the thermodynamic quantities at each time 
step \citep[Table~2 in][]{vanzanten:2011}. The radiative and advective cooling rates are prescribed jointly as a large-scale 
vertically homogeneous cooling rate profile of 2.5~K~day\textsuperscript{-1}. The sea surface temperature is set to 299.8~K, 
while the surface fluxes are computed interactively using a surface-layer bulk aerodynamic parameterization \citep[see][]{vanzanten:2011}. 
The geostrophic wind profiles are prescribed as time-invariant and equal to the initial wind profiles, and the background 
wind is set to $u=-5$~m~s\textsuperscript{-1} and $v=-4$~m~s\textsuperscript{-1}. Duration of the R-base simulation 
is 60~hours. 

To represent conditions over land, a second reference case (A-base) is set up based on the Atmospheric Radiation 
Measurement (ARM) campaign, as in \cite{brown:2002}. This case is forced by the averaged observed conditions at the 
Southern Great Plains (SGP) site on 21.~June~1997. The start of the simulation is set to 11:30~UTC (6:30~am by local time), 
a time before convection initiates, and is integrated over a single diurnal cycle until 02:00~UTC next day (21:00~pm by 
local time). The initial vertical profiles of the thermodynamics quantities are computed based on the averaged soundings 
from that day \citep[Fig.~1 in][]{brown:2002}. The wind direction did not change significantly during that day, so the 
initial wind profile is set to a constant wind of $u=10$~m~s\textsuperscript{-1} and $v=0$~m~s\textsuperscript{-1} 
at all levels. The geostrophic wind is also set to these values, while the background wind is set to $u=0$~m~s\textsuperscript{-1} 
and $v=7$~m~s\textsuperscript{-1}. At the surface, the turbulent heat fluxes are prescribed following \cite{brown:2002} 
(see their Fig.~3) and exhibit a strong diurnal cycle. Weak large-scale forcing tendencies due to horizontal advection of 
moisture and temperature as well as radiative cooling rates are prescribed following the diurnal cycle; however they have 
only a minor impact on the simulation. 

The two reference LES cases, R-base and A-base, are further modified to test the effects of  surface conditions, diurnal cycle 
and large-scale forcing on the cloud statistics (Table~\ref{cases}). For all LES cases the simulations are performed over a domain 
of 51.2 km $\times$ 51.2 km, with a horizontal grid spacing of 25~m and a vertical resolution of 25~m up to a height of 5~km 
(domain top). Five vertical grid levels are used as damping layers at the top of the domain.

\begin{figure*}[t]
	\centering
			\includegraphics[width=1\linewidth, clip=true, trim= 20mm  100mm 20mm 100mm]{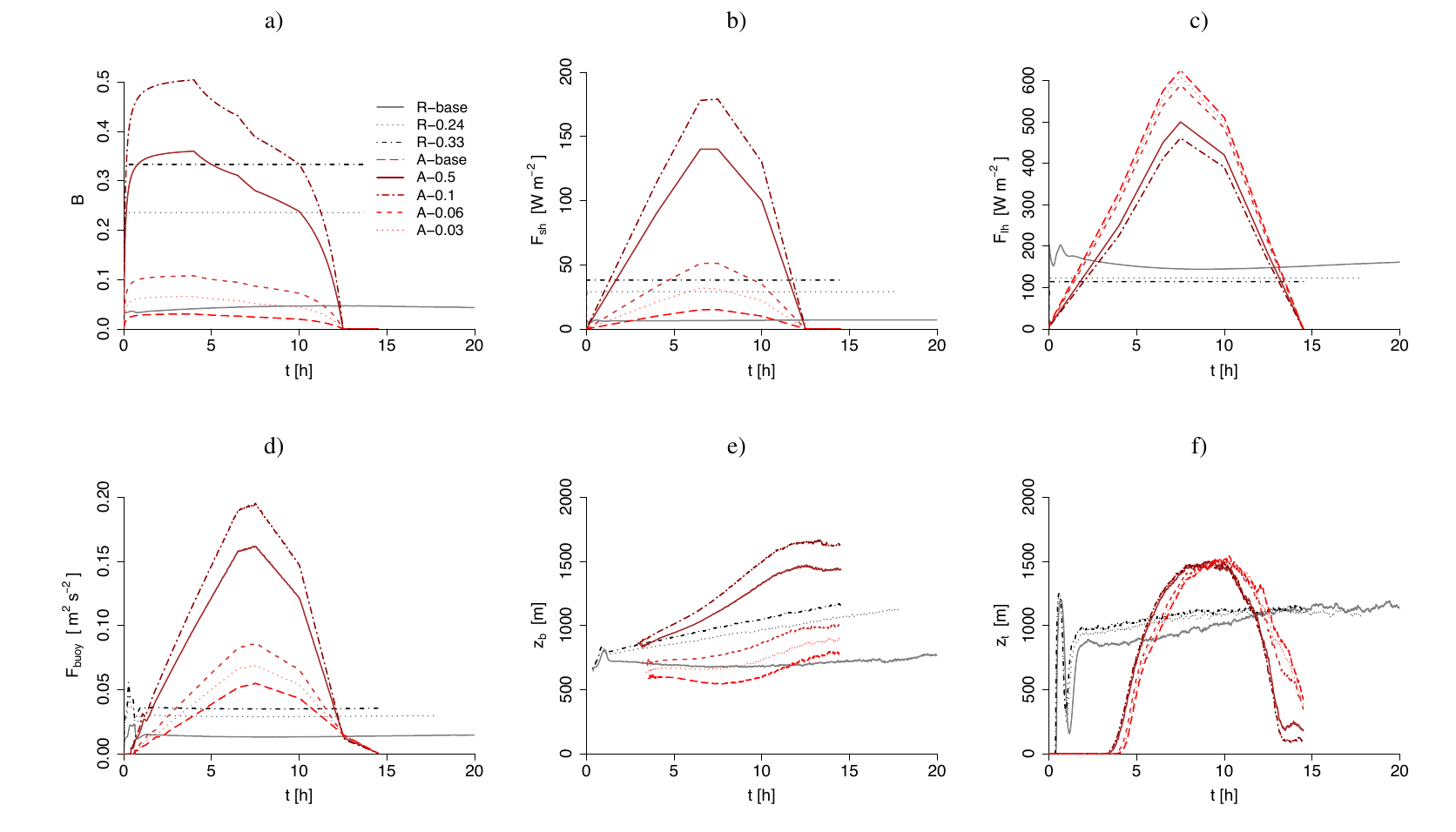}
	\caption{Time series of the surface forcing in the first group of eight LES cases from Table~\ref{cases}: a) Bowen ratio $B$, 
		b) surface sensible heat flux $F_{sh}$, c) surface latent heat flux $F_{lh}$, d) surface buoyancy flux $F_{buoy}$, and the 
		resulting e) cloud base $z_b$ and f) cloud top heights $z_t$. The difference between these simulations is set through 
		the Bowen ratio, which is indicated in the case abbreviations and line colours. Time from the start of the simulation 
		is shown on the x axis. }
	\label{sfc_for}
\end{figure*}

In the first group of simulations (R-base, R-0.24, R-0.33, A-base, A-0.5, A-0.1, A-0.06, and A-0.03; Figure~\ref{sfc_for}), 
we have prescribed a range of values of the ratio of the sensible to latent heat fluxes at the surface, the Bowen ratio, to both 
cases starting from $B=$~0.03 to $B=$~0.5. This range of values is selected because it encompasses the typical values of 
$B$ characteristic for the regions of the tropical oceans to midlatitude continental conditions. The purpose of these simulations 
is to investigate the hypothesis that the differences between the two reference cases come from different Bowen ratios. The 
average Bowen ratio in R-base is around 0.03 and is approximately constant, while in A-base the starting value of B is around 
0.3, and it decreases slightly over the diurnal cycle (Figure~\ref{sfc_for}a). The two simulations based on RICO, R-0.24 and 
R-0.33, are set up by fixing the surface heat fluxes instead of the fixed SST. The total heat flux magnitude is kept equal to the 
reference RICO case, but the ratio of sensible to latent heat flux is changed to result in the wanted B value, 0.24 in the first and 
0.33 in the second case. In the ARM-based cases (A-0.5, A-0.1, A-0.06, and A-0.03) the total surface heat fluxes are kept the 
same, but the ratio of sensible to latent heat flux is changed to result in the targeted B values of 0.5, 0.1, 0.06, and 0.03. These 
new B values are set at the beginning of the diurnal cycle, and are decreasing over the cycle at the same rate as in A-base 
(see Figure~\ref{sfc_for}a). Note that the total surface heat flux in the RICO-based cases is in average more than twice lower 
than the total surface heat flux in the ARM-based cases (Table~\ref{cases}). By comparing the maximum values of the total 
surface heat flux or of the buoyancy flux near the peak of the diurnal cycle (Figs.~\ref{sfc_for}b-d), the difference between 
the two reference cases is even up to four times. 

As expected, the mean thermodynamic state of the subcloud layer is affected by the changes in the Bowen ratio. Increase of the 
Bowen ratio from 0.03 to 0.33 in the RICO-based cases causes an increase of the liquid water potential temperature by 1~K, and 
a decrease in the total water mixing ratio by 1~g/kg, as averaged over a 500~m thick layer starting from the surface. In the ARM 
case, a decrease of the Bowen ratio from 0.33 in A-base to 0.03 in A-0.03 causes a decrease of the liquid water potential temperature 
by 2~K, and an increase of the total water mixing ratio by 2~g/kg, averaged over a 500~m thick layer at the surface. Clearly, all these 
test cases have a different thermodynamic state in the boundary layer, even though the Bowen ratios might have the same values. 

The depth of the subcloud layer is controlled by the surface buoyancy flux $F_{buoy}$ \citep{stevens:2007} with the higher cloud 
base heights in the simulations with higher surface buoyancy fluxes (Figs.~\ref{sfc_for}d and \ref{sfc_for}e). The rate of growth 
of the subcloud layer is also influenced by $B$ and it is higher 
in the cases with higher $B$  \citep[Figure~\ref{sfc_for}e, see also ][]{schrieber:1996}. Convective clouds are initiated sooner for 
the higher values of $B$ (Figure~\ref{sfc_for}f). Except for the R-base case where the surface fluxes are not fixed, the top of the 
cloud layer does not seem to be significantly influenced by the changes in $B$ or $F_{buoy}$  (Figure~\ref{sfc_for}f). This indicates 
that the processes in the cloud layer are to some extent detached from the surface forcing.

\begin{figure*}[t]
	\centering
				\includegraphics[width=1\linewidth, clip=true, trim= 20mm  100mm 20mm 100mm]{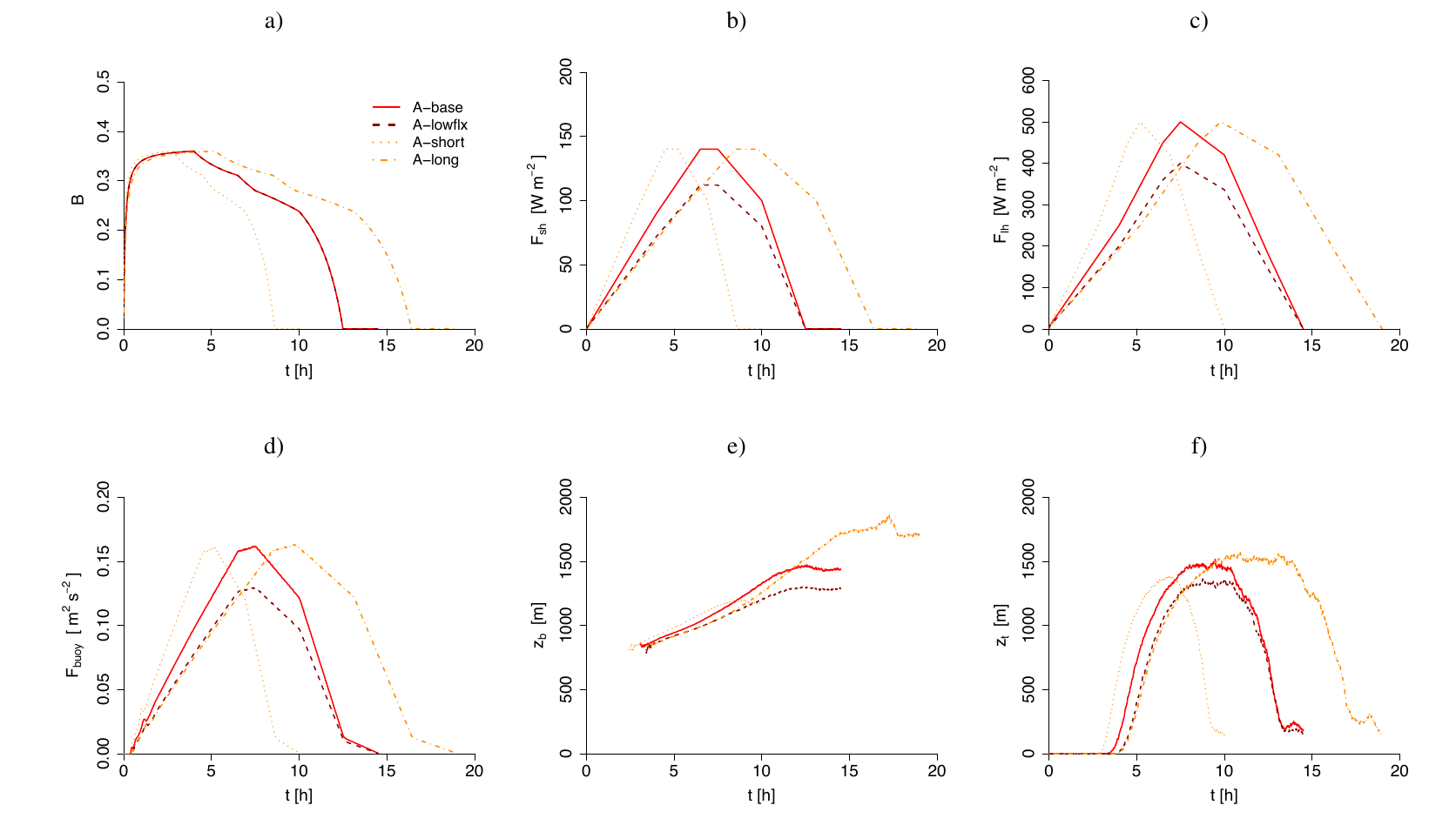}
	\caption{Time series of the surface forcing in the second group of the LES cases based on ARM: a) Bowen ratio $B$, b) surface 
		sensible heat flux $F_{sh}$, c) surface latent heat flux $F_{lh}$, d) surface buoyancy flux $F_{buoy}$, and the resulting 
		e) cloud base $z_b$ and f) cloud top heights $z_t$. The difference between these simulations is set through the period of the 
		large-scale forcing (A-short and A-long) and through the total surface heat flux magnitude (A-lowflx).}
	\label{period_ls}
\end{figure*}

In the second group of simulations (A-lowflx, A-short, A-long; Figure~\ref{period_ls}), we have kept the Bowen ratio to its assigned 
values, but changed other key aspects of the forcing that are distinct between the two reference cases. The effect of the diurnal cycle 
in ARM is tested by shortening it by 1/3 (A-short), or by prolonging it by 1/3 (A-long), by applying these changes to the cycle period 
of the surface fluxes (see Figs.~\ref{period_ls}b,c) and the large-scale forcing tendencies. The effect of the value of the total surface 
heat flux is tested by reducing it by 20\% in ARM (A-lowflx). As can be seen in Figure~\ref{period_ls}e, the rate of the growth of the 
cloud base height is not affected by these changes. However, if there is more time for the cloudy boundary layer to develop, as in A-long, 
a higher cloud base height is reached. The cloud layer deepens further either with an increase in the forcing period or with stronger 
total surface heat fluxes, although the differences are only around 100~m (Fig.~\ref{period_ls}f).

\subsection*{Cloud tracking}

The cloud tracking algorithm developed by \cite{heus-seifert:2013} is applied to the simulated cloud fields in post-processing of the 
LES simulations. In the tracking algorithm, clouds are identified as the adjacent grid points that hold the liquid water path exceeding 
a threshold value of 5~gm\textsuperscript{-2}. In that way, the identified cloud area is a projection of a cloud from all vertical levels 
that can be tracked through space and time. Using the temporal resolution of one minute, cloud areas, vertical velocities and cloud 
lifetimes are recorded for each cloud in the simulation. A cloud splitting algorithm is then used to separate and track the individual 
cloud elements that form the multicore clouds or the merged cloud clusters. These cloud elements are defined as holding a buoyant 
core with the maximum incloud virtual potential temperature $\theta_v$ excess larger than a chosen threshold of 0.5~K. More details 
and validation of the tracking method are provided in \cite{heus-seifert:2013}. 

To develop a cloud parameterization based on the mass flux approach, the cloud mass flux has to be estimated near the cloud-base 
level. For this reason, we have developed a secondary tracking routine, as in \cite{sakradzija:2015}, in which we record the area that 
every cloud occupies at the level that lies 100 m above the lifting condensation level (LCL). We define this area as the area that contains 
all the points with liquid water content greater than zero. 

\section{Cumulus cloud population statistics}

The upward flux of mass through cloud base of the i-th cloud is defined as 
\begin{equation}
m_i = \rho a_i w_i
\end{equation}
where $a_i$ is the area [m\textsuperscript{2}] occupied by points holding liquid water at a level 100~m above LCL and 
$w_i$~[m~s\textsuperscript{-1}] is the vertical velocity averaged over the area $a_i$. To compute the distribution of the 
cloud mass flux, $p(m)$, we average $m_i$ over the lifetime of each cloud. Similar results can nevertheless be obtained 
by looking at the instantaneous values. The choice of computing the lifetime averaged mass fluxes comes from the 
possibility to reconstruct cloud lifecycles for the purpose of a parameterization, as in \cite{sakradzija:2015}. 

\begin{figure}[t]
	\centering
	\begin{tabular}{c}
		{\includegraphics[width=0.4\linewidth, clip=true, trim= 0mm  0mm 0mm 5mm]{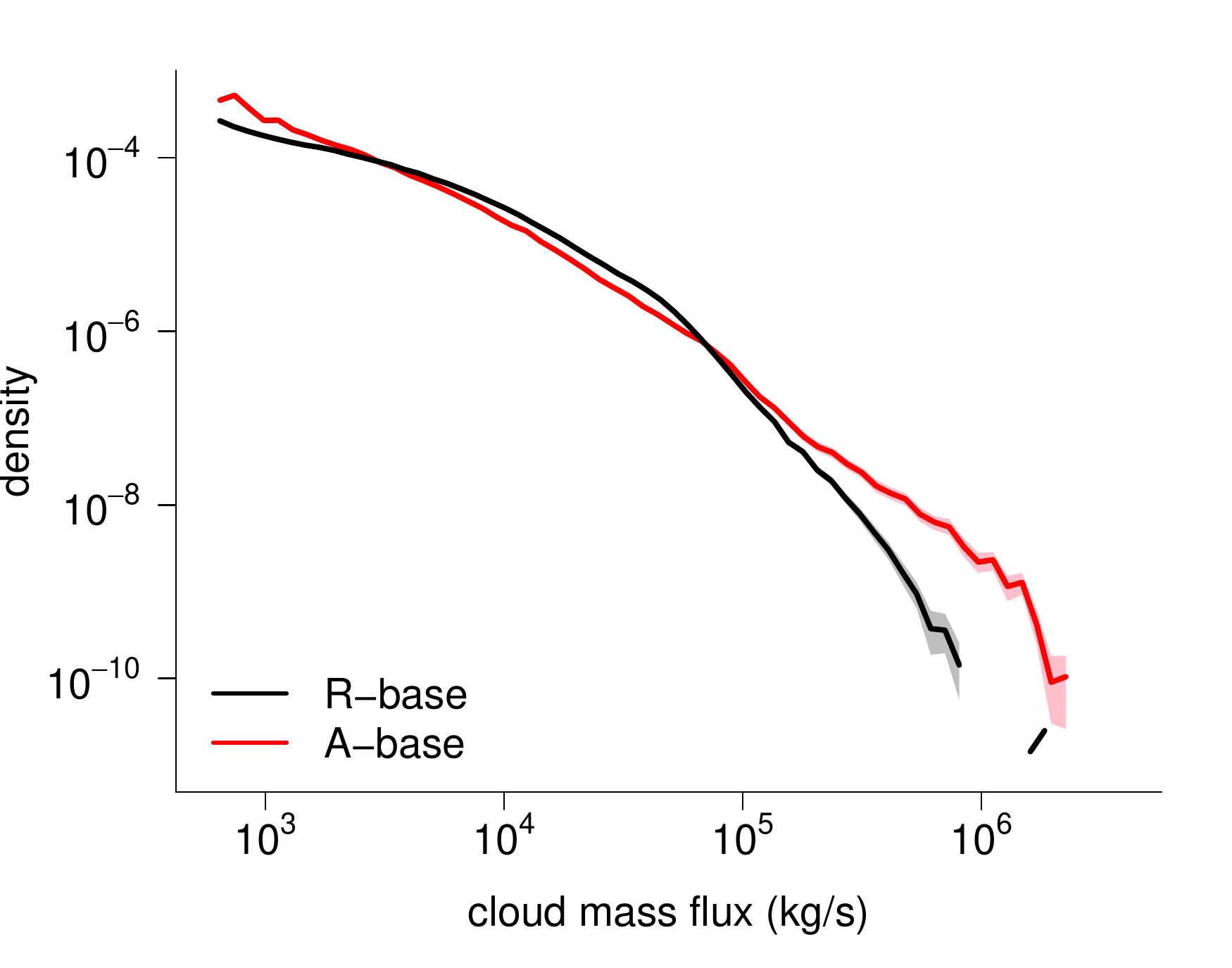}} \\
	\end{tabular}
	\caption{ The probability density distribution of the lifetime average cloud base mass fluxes. Cloud are sampled from the 
		6th to 22nd hour
		from the simulation start in the R-base case, and from the 9th to 12th hour after the simulation start in the A-base case. 
		Clouds with mass flux values lower than 600~kg/s are discarded from the plot to remove possible numerical noise, 
		since those are mostly the clouds that cover only a single grid cell. 95~\% confidence bands are plotted as shaded areas. }
	\label{base-mf}
\end{figure}

The distribution of cloud base mass fluxes is calculated for the two reference cases, RICO and ARM (Fig.~\ref{base-mf}). 
The probability density distribution is computed using the generic R function \textit{hist} \citep{R}. The width of the bins 
used to compute the probability density of mass fluxes is logarithmically increasing with higher mass flux values. The 
sampling period in RICO is from the 6th to the 22th hour of simulation, while in the ARM case clouds are sampled from 
the 6th (17:30~UTC) to the 12th (23:30~UTC) simulation hour. Only those clouds that were initialized during the sampling 
period are included in the calculation. Clouds that lasted longer than this sampling time period are followed beyond the 
time limit to finalize their lifecycles. The sample size of the lifetime average cloud base mass flux distribution is 317~014 
clouds in the RICO case, and 120~292 clouds in the ARM case. 

The two reference LES cases exhibit distinct horizontal and vertical extents of the clouds, number of clouds and their spacing, 
due to different initial conditions, surface and large-scale forcing. The mass flux distributions corresponding to these two 
reference cases have different shapes and they cover different ranges of the mass flux values (Fig.~\ref{base-mf}). The distribution 
of the cloud base mass flux in the ARM case shows a straight line shape on a log-log plot, similar to a power-law distribution 
over a range of three orders of magnitude.  In contrast, the distribution in the RICO case shows a more concave shape. In previous 
literature on the cloud size distribution, such type of a concave shape has often been identified as a double power-law distribution 
with two distinct slopes and a scale-break point at the intermediate cloud size 
\citep{cahalan-joseph:1989, segunpta:1990, nair:1998, benner-curry:1998, neggersetal:2003, trivej-stevens:2010, heus-seifert:2013}. 
To make a parallel to these studies, we identify the scale-break in the mass flux distribution of the R-base case at a value of 
the cloud base mass flux close to $1\cdot10$\textsuperscript{5}~kgs\textsuperscript{-1} (Fig.~\ref{base-mf}). Based on the qualitative comparison 
of the mass-flux distributions of the R-base and A-base case, we conclude that there is no universality in the distribution slopes 
on a log-log plot (Fig.~\ref{base-mf}). As we will show in section~4c, the slope of the mass flux distribution changes with the 
change of a control parameter of the simulations. 
 
The sampling variability of the mass-flux distributions is very low in both reference cases except near the end of the right 
tails of the distributions (Fig.~\ref{base-mf}), which is a sign of a limited sample size of the largest possible cloud mass flux values. 
This portion of the distribution tail has higher sampling variability based on the 95~\% confidence intervals computed for each 
distribution bin (shaded areas in Fig.~\ref{base-mf}). The confidence intervals were calculated using a bootstrap method with 
replacement using 1000 random samples. 

\begin{figure}[t]
	\centering
		{\includegraphics[width=1\linewidth, clip=true, trim= 0mm  90mm 0mm 90mm]{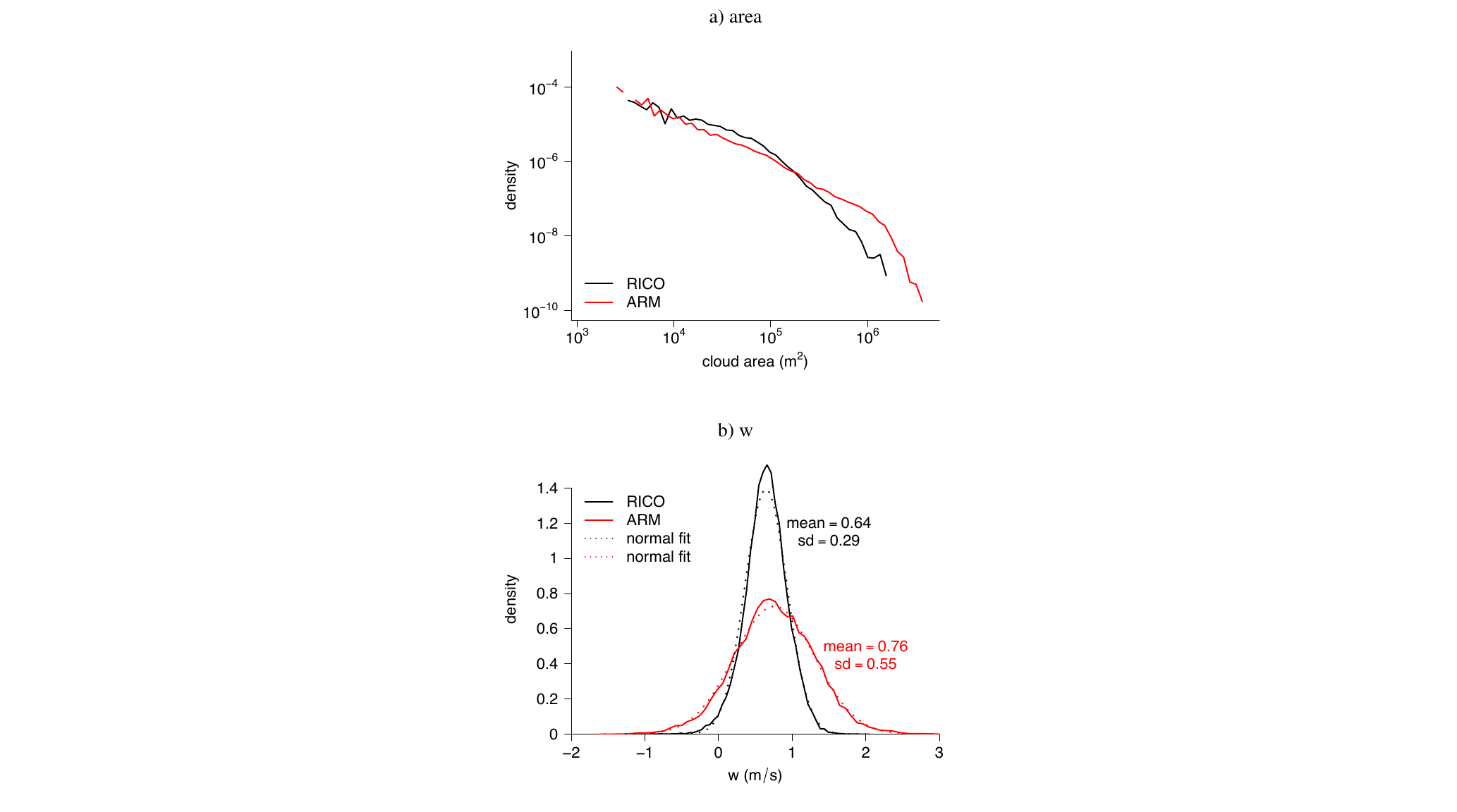}} \\
	\caption{The probability density distribution of a) lifetime averaged cloud base areas and b) vertical velocity through cloud base. 
		As in the previous figure, clouds with mass flux values lower than 600~kg/s are discarded from the plot to remove possible 
		numerical noise.}
	\label{area-w}
\end{figure}

As a key contributor to the cloud base mass flux, the cloud area $a_c$ is distributed qualitatively similarly to the distribution 
of the mass flux (Fig.~\ref{area-w}a). The difference between the two reference LES cases shows similar characteristics as for 
the two mass flux distributions. So, the knowledge about the physical mechanism that shapes $p(m)$ might also be sufficient 
to describe $p(a_c)$. The cloud area distribution of the A-base case shows a power-law-like shape with a scale-break around  
the value of 10\textsuperscript{6}~m\textsuperscript{2}. The scale break in the ARM-base case is located at a scale 
an order of magnitude larger than the one of the R-base case. These two cloud 
area distributions are actually very similar to the two typical cloud size distributions observed over land and over ocean as 
derived from the Landsat images in \cite{segunpta:1990}, their Fig.~4. A similar change in the distribution behaviour for the 
largest cloud areas is observed in the radar echo areas distribution in \cite{trivej-stevens:2010}. Different statistics of the large 
echoes compared to a power-law behaviour of the small echoes may be controlled by the meteorological environment. In particular, 
the existence of an inversion layer topping the cloud layer limits the growth of clouds beyond a certain size, which can be connected 
to the observed break in the scaling \citep{trivej-stevens:2010}. Strong subsidence inversions over the tropical oceans might 
explain the position of the scale-break at the lower values than what is observed at midlatitudes \citep[see also ][]{wood-field:2011}.

The distribution of vertical velocity of individual clouds is approximately symmetric and can be well fitted using a normal 
distribution, as illustrated in Fig.~\ref{area-w}b. The average vertical velocity per cloud is $\langle w \rangle=0.64$ 
in the RICO case and $\langle w \rangle=0.76$ in the ARM case. Compared to the RICO case, in the ARM case the variance 
of $w$ is significantly higher and some clouds can gain velocities larger than 2~m/s. This result is in line with the findings 
of \cite{xu-randal:2001}, albeit for deep convection, where the most significant differences in the updraft intensities 
between tropical oceanic and midlatitude continental convection were found in the strongest 10\% of the updrafts, not in 
the median values. The correlation between vertical velocity $w_i$ of individual clouds and their mass fluxes $m_i$ 
is very low (not shown here). This is the reason for the similarity between $p(m)$ and $p(a_c)$, while $p(w)$ belongs 
to a different family of distributions. 

Why are the two reference population distributions different? Is the distribution shape changing under the influence of 
the large-scale forcing or of the surface conditions? We address these questions in the following section.

\section{The three hypotheses}

The main differences between the two reference LES cases are in the existence of a strong diurnal cycle over land, strong 
self-organization of clouds over ocean and in the magnitude and partitioning of the surface turbulent heat fluxes (Table~\ref{cases}). 
Other aspects of the large-scale forcing are as well different between the two reference cases. However, we rule out those 
differences as a cause of the different distribution shapes because it was hypothesised and shown in previous studies that 
the intensity of the convective updrafts was insensitive to changes in the large-scale forcing \citep[e.g.][]{robe-emanuel:1996, cohen-craig:2006,pc:2008}. 
Based on these facts, we propose the three hypotheses that might explain the divergence of the mass flux distribution 
between the two reference LES cases:
\begin{enumerate}[a.]
	\item  diurnal cycle of convection determines the distribution $p(m)$, 
	\item  convective self-organization determines the distribution $p(m)$, and
	\item  surface fluxes determine the distribution $p(m)$. 
\end{enumerate}
In the following, we test the three hypotheses by analysing all eleven LES cases (Table~\ref{cases}).

\subsection{The first hypothesis: diurnal cycle of convection}

\begin{figure}[h]
	\centering
	\begin{tabular}{c}
		{\includegraphics[width=0.4\linewidth, clip=true, trim= 0mm  0mm 0mm 0mm]{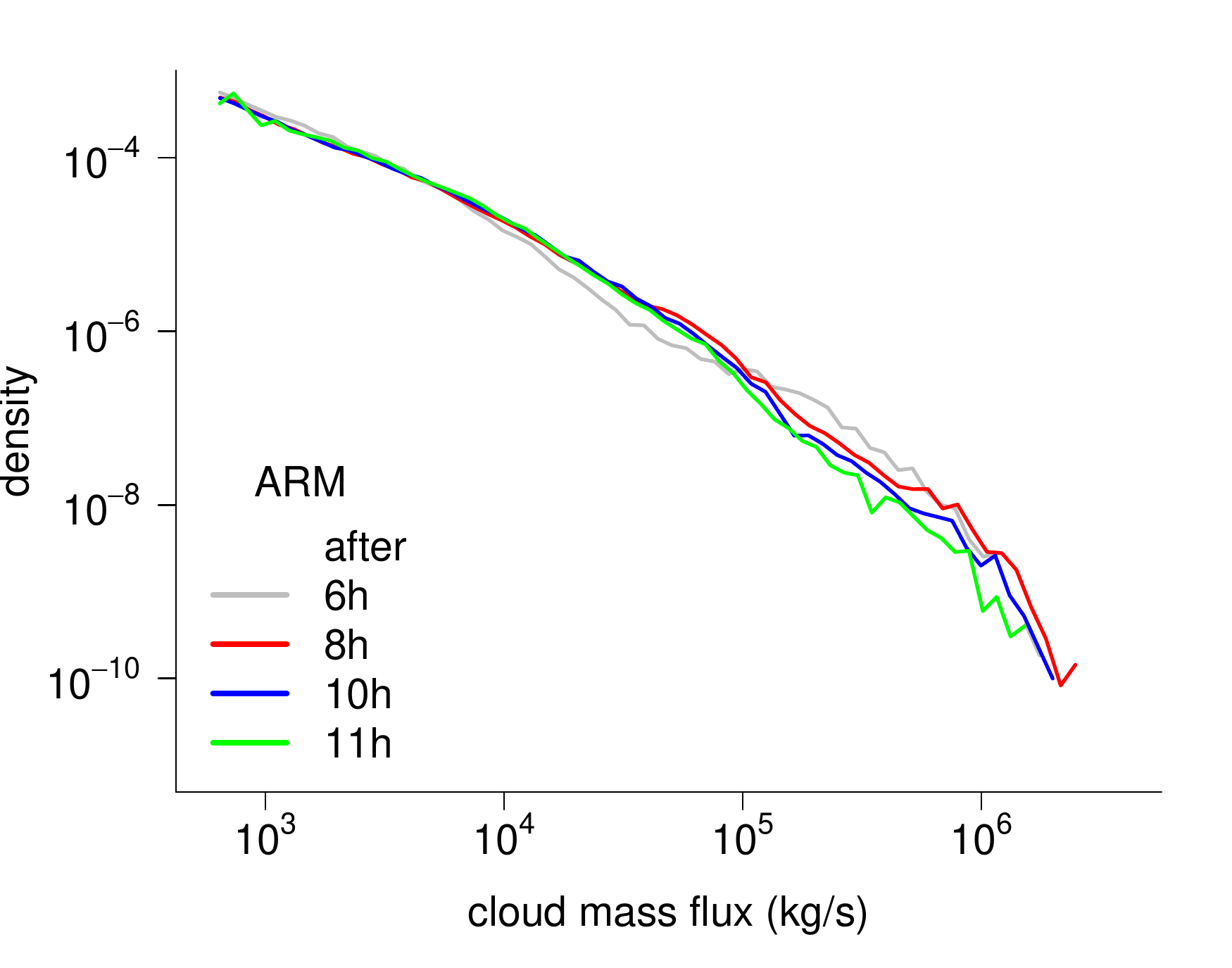}} \\
	\end{tabular}
	\caption{The probability density distribution of the lifetime average cloud base mass flux sampled over time frames of 
		one hour duration over the diurnal cycle of the A-base case, starting after 6, 8, 10 and 11 hours of simulation 
		(at 17:30~UTC, 19:30~UTC, 21:30~UTC, and 22:30~UTC).}
	\label{diur_cyc}
\end{figure}

Here we test if changes in the forcing associated with the convective diurnal cycle might be responsible for the different 
shapes of $p(m)$ in the two reference cases. We sample the clouds that emerge in the ARM case during four time frames 
of one hour duration, taken at different stages of the diurnal cycle, starting at 17:30~UTC. The distribution of cloud base 
mass flux in all four time frames is shown in Fig.~\ref{diur_cyc}. It is clear that there is no significant change in $p(m)$ 
over the diurnal cycle of the ARM case, i.e. the distribution $p(m)$ is stationary. 

\begin{figure}[h]
	\centering
	\begin{tabular}{c}
		{\includegraphics[width=0.4\linewidth, clip=true, trim= 0mm  0mm 0mm 0mm]{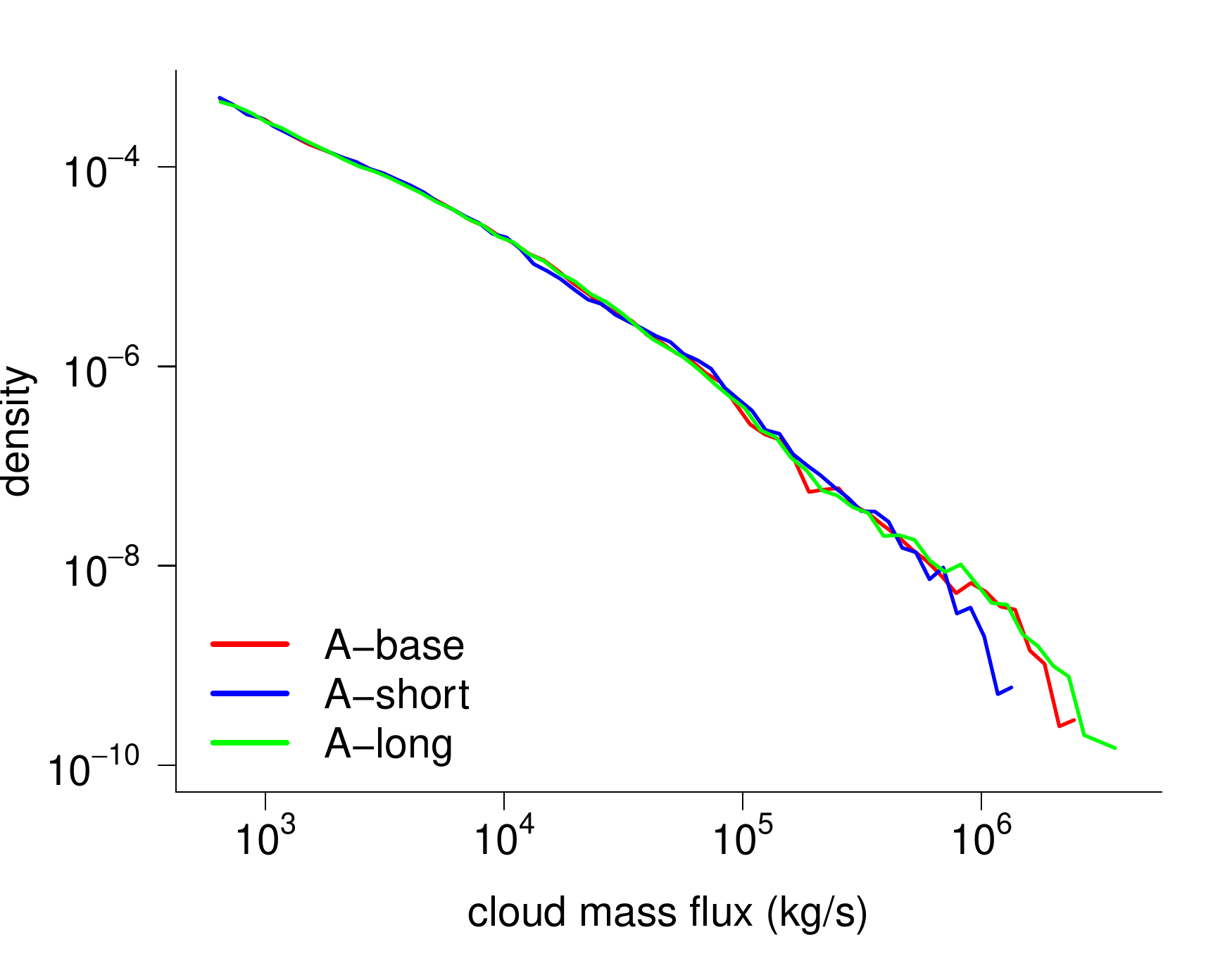}} \\
	\end{tabular}
	\caption{The probability density distribution of the lifetime average cloud base mass flux over the cases with different diurnal cycle 
		periods, based on the ARM case. }
	\label{diur_cyc2}
\end{figure}

Another property of the diurnal cycle that might influence $p(m)$ is the period of the diurnal cycle. Shorter or longer 
diurnal cycles imply faster or slower temporal changes in the forcing. With faster changes, clouds might have less time 
to develop undisturbed, so their sizes and mass fluxes might be lower. Or, with slower changes in the forcing, larger 
clouds might result. To test this, we investigate the results of the simulations A-short and A-long. A time frame of one 
hour duration is taken around the peak of the diurnal cycle, after 9, 7 and 11 hours from simulation start in A-base, 
A-short, and A-long, respectively, and $p(m)$ is examined (Fig.~\ref{diur_cyc2}). There is again no significant difference 
among the simulations, except near the right tail of the distribution, where the A-short case shows a faster drop-off than 
the other two cases. This means that the largest possible clouds cannot develop in the ARM case if the period of the forcing 
is too short. Overall, there is nevertheless no change in the distribution shape, and the slope of the line stays similar across 
the three cases. 
The results of these experiments demonstrate that changes of the forcing over a diurnal cycle do not shape the 
distribution of the cloud base mass flux.

\subsection{The second hypothesis: convective self-organization}

In this section we test how the spatial correlations during the organized phase of the RICO case influence the cloud 
base mass flux distribution. 
Organization of convective clouds into clusters, lines, or arcs could influence $p(m)$ by affecting the size and intensity 
of  individual cloud elements. Here, it is important to note that the cloud tracking routine identifies the cloud entities 
that form the cloud clusters, and performs splitting so that every element can be followed separately even when two 
cloud elements have merged. In contrast, past studies have investigated the distributions of merged cloud clusters 
and suggested self-organization as a mechanism for creating power-laws \citep{scheufele:2014}. 

\begin{figure}[h]
	\centering
			\includegraphics[width=1\linewidth, clip=true, trim= 10mm  90mm 10mm 90mm]{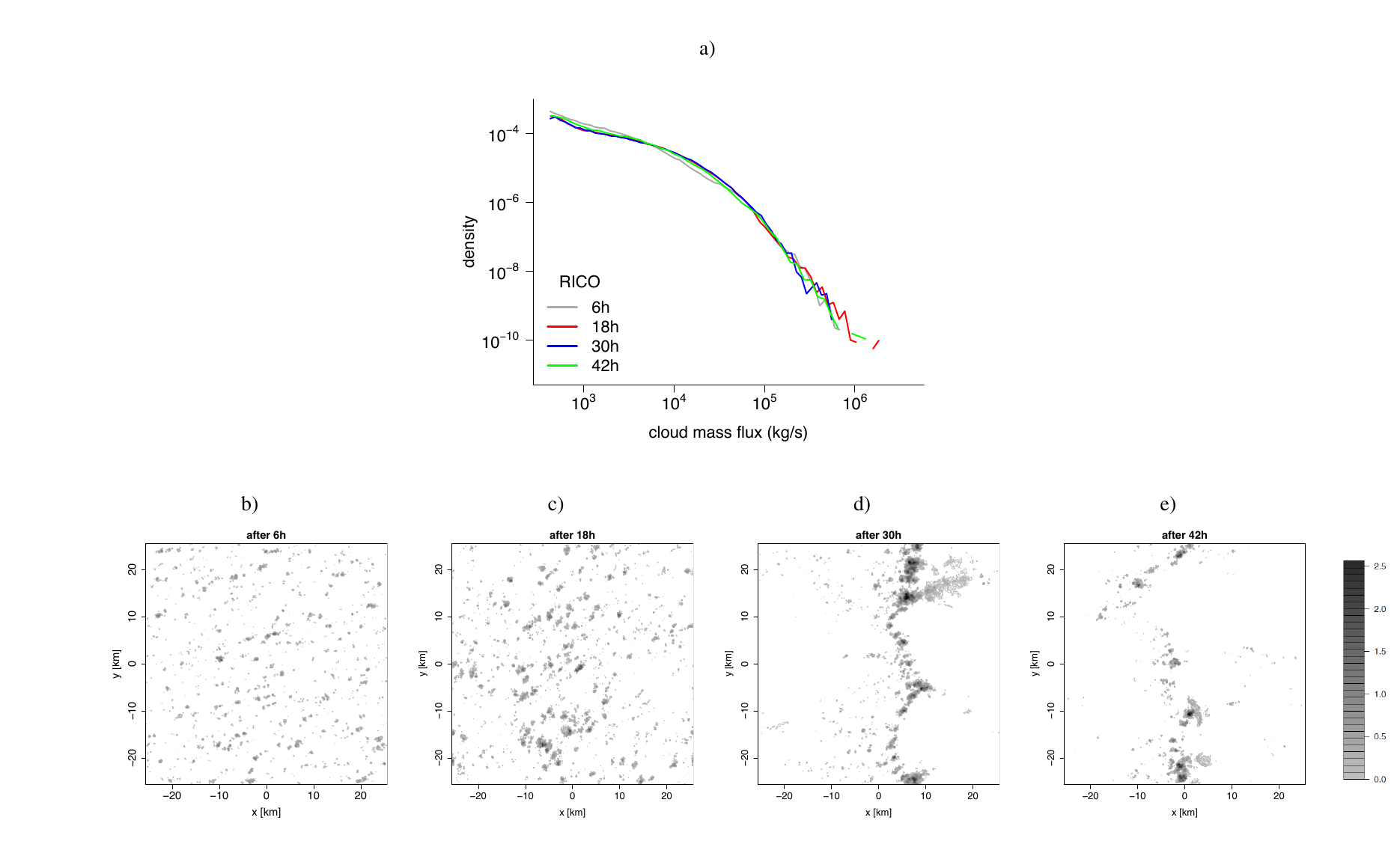}
	\caption{The probability density distribution of the lifetime average cloud base mass flux in the R-base case, over 
		different stages of cloud organization. The corresponding horizontal spatial distribution of the cloud field is visualized 
		in (b-e) using the liquid water path [g~m\textsuperscript{-2}]. }
	\label{org}
\end{figure}

We choose the R-base case to test the effects of cloud organization on $p(m)$ because this convective case is 
strongly organized after one day of simulation (Fig.~\ref{org}). Starting from a randomly distributed field of 
clouds and looking into the time frames with different stages and forms of organization, we plot $p(m)$ in Fig.~\ref{org}. 
We find no evidence that self-organization of clouds has an effect on $p(m)$ because the overall distribution 
shape stays the same in spite of organization. Hence, the different degrees of organization between RICO and 
ARM cannot explain the differences in $p(m)$. 

Even though self-organization is not responsible for the final shape of the distribution, it is a process that can 
produce longer tails in the cloud distributions, if the cloud splitting is not performed and cloud clusters are 
sampled to compute $p(m)$ \citep{scheufele:2014}. Fig.~\ref{org} indicates that this dependency vanishes if 
individual cloud elements are considered. 

\subsection{The third hypothesis: surface heat fluxes}

The two reference cases have very different surface conditions, one is set over the ocean, while the other one is 
set over land, so the magnitudes of the surface heat fluxes differ by up to a factor four between the cases (see  
Fig.~\ref{sfc_for}). We investigate here the dependency of the distribution shape on the surface turbulent heat 
fluxes, which drive the boundary layer convective updrafts that ultimately form cumulus clouds at the top of the 
subcloud layer. We test the magnitude of the fluxes and their partitioning at the surface. 

\subsubsection{The magnitude of the surface heat fluxes}

\begin{figure}[th]
	\centering
	\begin{tabular}{c}
		{\includegraphics[width=0.4\linewidth, clip=true, trim= 0mm  0mm 0mm 0mm]{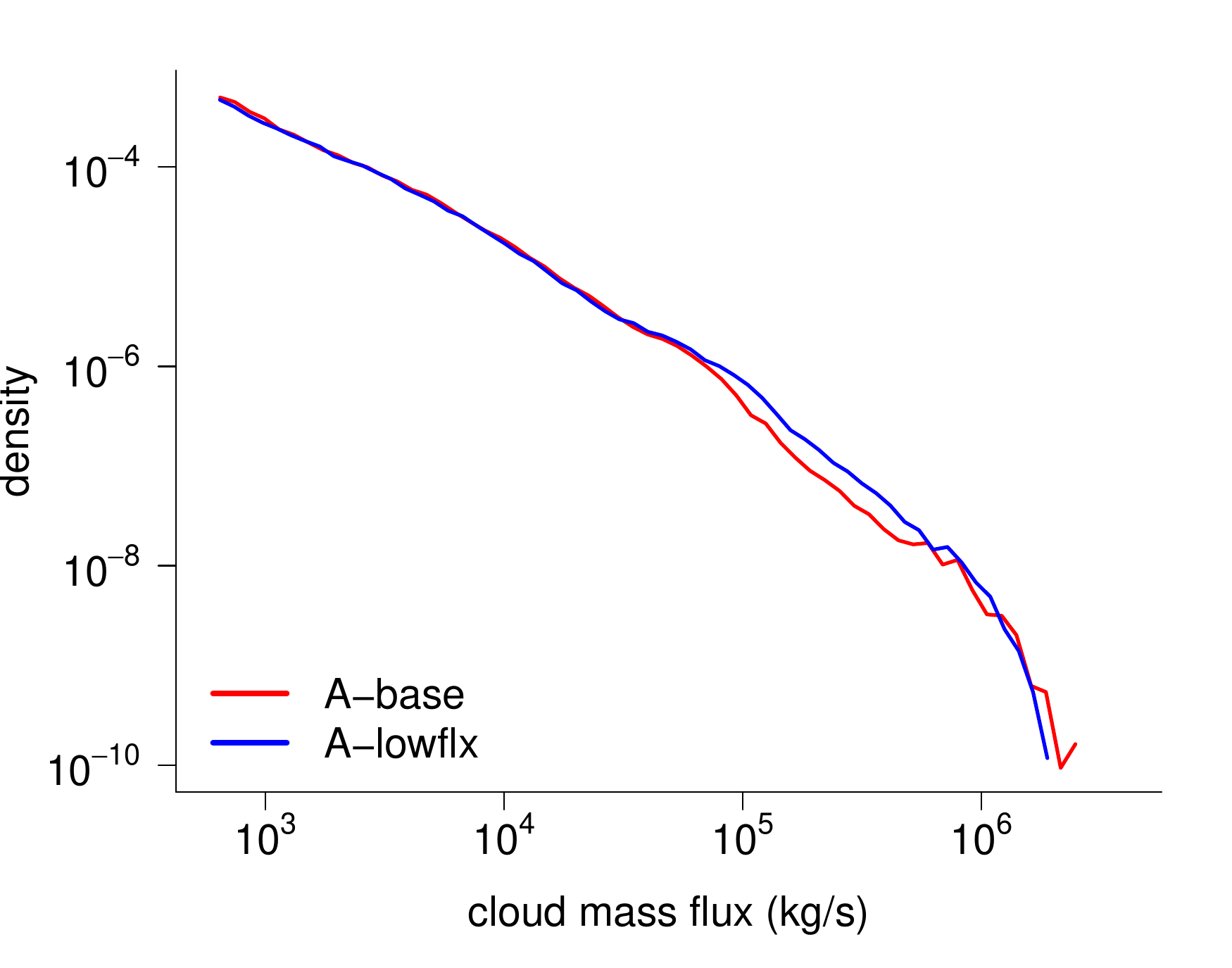}} \\
	\end{tabular}
	\caption{The probability density distribution of the lifetime average cloud base mass flux in the A-lowflx case 
		compared to the distribution in the A-base case. }
	\label{Fin}
\end{figure}

We have already concluded in the previous section for the ARM case that $p(m)$ does not change over a single 
diurnal cycle (Fig.~\ref{diur_cyc}). From this conclusion it also follows that $p(m)$ is not sensitive to the surface 
flux magnitude. To further prove this, we perform one additional test (A-lowflx) in which the total surface 
turbulent heat flux is lowered by 20~\% (Fig.~\ref{Fin}). There is no significant difference between the two 
distributions. The A-lowflx case can simply be considered as another realization of the same shallow cloud 
ensemble of the A-base case. 

\subsubsection{The ratio of the surface heat fluxes, $B$}

\begin{figure}[t]
	\centering
		{\includegraphics[width=1\linewidth, clip=true, trim= 0mm  90mm 0mm 80mm]{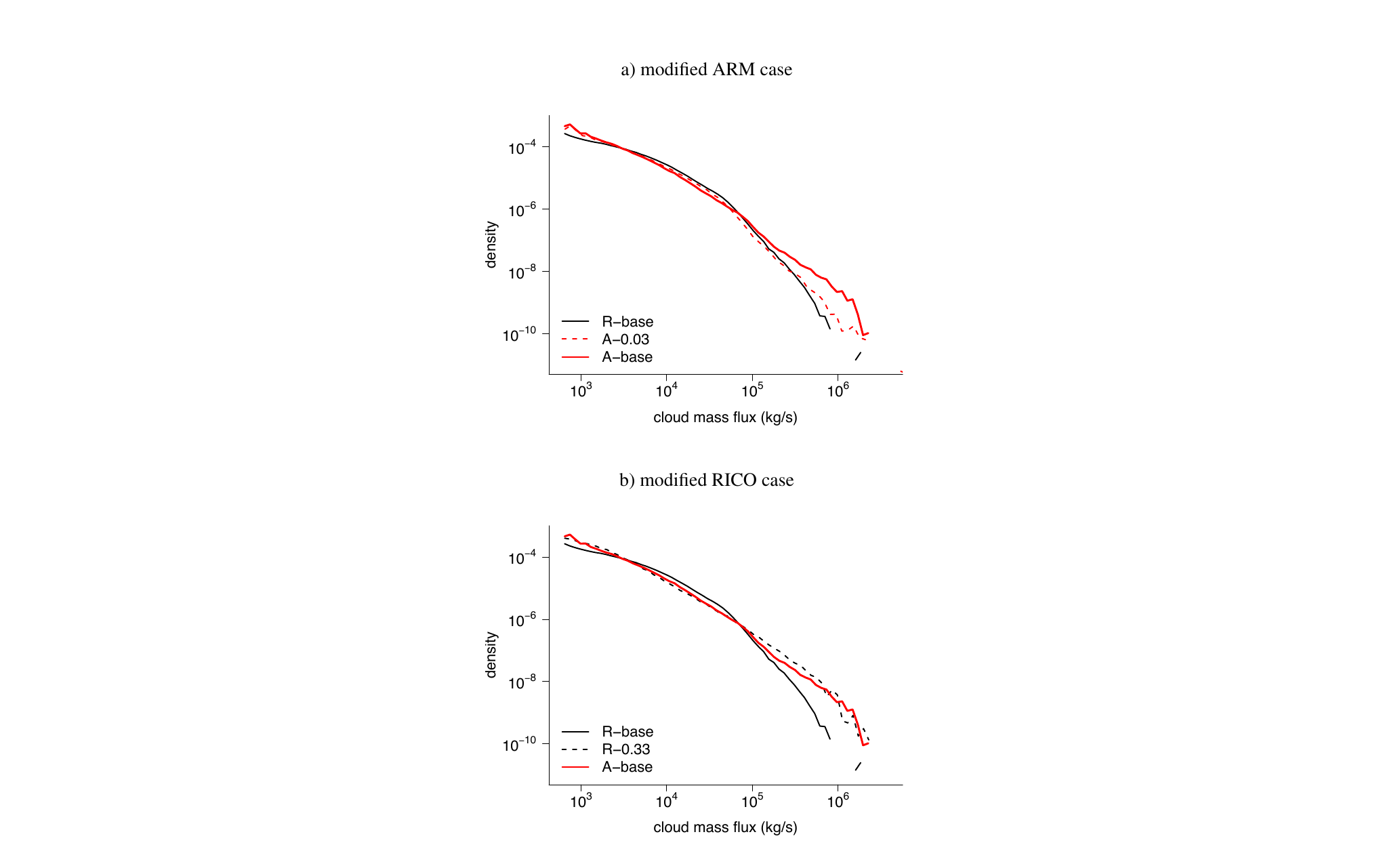}} \\
	\caption{ Reproduction of the distribution shape of the lifetime average cloud base mass flux of a) the R-base case 
		by altering the Bowen ratio of the ARM case to $B=0.03$,  
		and b) of the A-base case by altering the RICO case Bowen ratio to $B=0.33$. }
	\label{B}
\end{figure}

The ratio of the sensible and latent heat fluxes at the surface, the Bowen ratio $B$, is the main parameter that 
characterizes the two surface types, ocean and land. Though the total surface flux magnitude has no effect on 
$p(m)$, the partitioning of this flux into sensible and latent heating might have an effect. Note that the Bowen 
ratio does not change much over the diurnal cycle in ARM. We thus turn our attention to the sensitivity experiments 
using different Bowen ratios (Fig.~\ref{B}).

By changing only the ratio of the surface fluxes and leaving their magnitudes unchanged, the shape of the mass 
flux distribution can be altered. More importantly, by setting the RICO Bowen ratio in the ARM set-up (A-0.03), 
the mass flux distribution of the RICO case is recovered (Fig.~\ref{B}a). Likewise, by setting the ARM Bowen ratio 
in the RICO set-up (R-0.33), the mass flux distribution of the ARM case is recovered (Fig.~\ref{B}b). Thus, it is 
evident that the ratio of the surface fluxes and not their magnitudes shapes the mass flux distribution.

\subsubsection{The two modes of the cloud distribution}

The final shape of $p(m)$ is a result of the superposition of the distribution modes associated with cloud groups 
of different subtypes: active, forced and passive clouds \citep[see the classification of][] {stull:1985}. We examine 
the dependency of these modes on the Bowen ratio separately. Here we simplify the classification of clouds into 
buoyant (active) and non-buoyant (passive) clouds, as in \cite{sakradzija:2015}.  Forced clouds fall into the "passive" 
non-buoyant cloud group owing to this simplification. Clouds are classified as active buoyant clouds if the excess of the 
vertically integrated virtual potential temperature within clouds, $\theta_{v,up}-\overline{\theta_v}$, is larger than a 
threshold. The threshold is set to 0.5~K, except in a case where this leads to a too small statistical sample, as in 
R-0.33. In the latter case, the threshold is set to 0.4~K.  

\begin{figure}[t]
	\centering
		{\includegraphics[width=1\linewidth, clip=true, trim= 0mm  80mm 0mm 80mm]{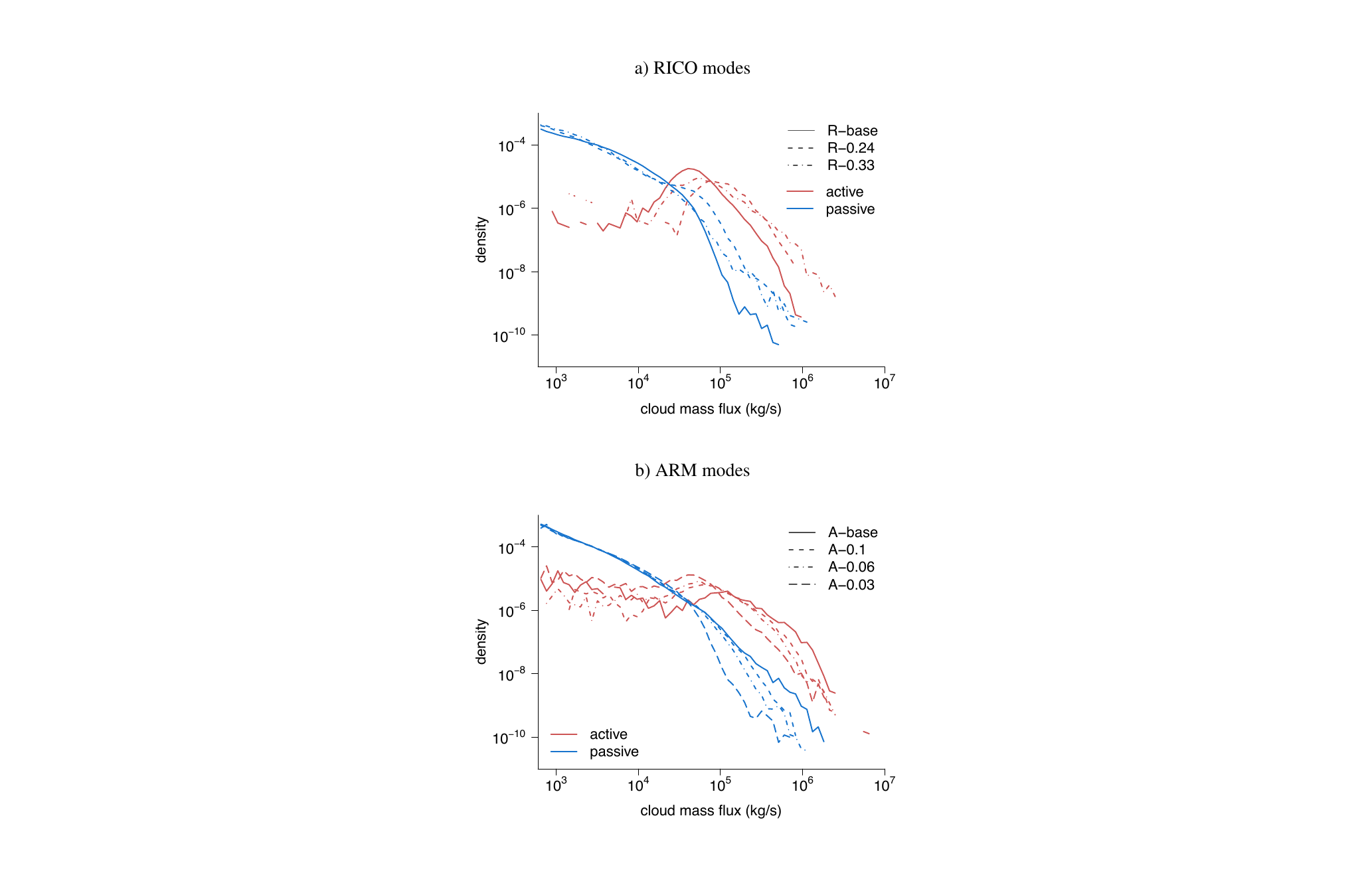}} \\
	\caption{The lifetime average cloud base mass flux distribution of active and passive cloud modes for different Bowen ratios: a) RICO based cases, and 
		b) ARM based cases. Full lines correspond to the reference cases, while dashed lines correspond to the cases with changed $B$.}
	\label{act_pas_B}
\end{figure}

In the RICO-base case (Fig.~\ref{act_pas_B}a), the cloud distribution shows shorter tails in both modes, and lower 
mass fluxes in average compared to the A-base case (full lines in Fig.~\ref{act_pas_B}b).  With increasing Bowen 
ratio, the active cloud modes shift towards higher mass flux values, while the slopes of the two modes become 
less steep (Fig.~\ref{act_pas_B}). Through the control on the range of mass flux values that individual modes of 
$p(m)$ can take, and by setting the slope of the modes, the Bowen ratio ends up determining the average mass 
flux per cloud $\langle m \rangle$ in both distribution modes. 

This fact might explain why different power-law slopes are documented in different observational studies of cloud population 
\citep[see Table 1 in][]{zhao-digirolamo:2007}. The slopes of the observed cloud size distributions in the midlatitude regions 
have lower values than the slopes in the tropics \citep[see][]{wood-field:2011}. These characteristics of the observed cloud size 
distribution correspond to the control that $B$ imposes on the slopes that we observe in the RICO and the ARM cloud-base mass 
flux distributions (Fig.~\ref{act_pas_B}). Higher values of $B$ in midlatitudes produce lower slopes compared to the 
higher slopes that are produced as a result of low $B$ in the tropics. 

\section{The Bowen ratio indirectly sets the average mass flux per cloud }

To understand the link between $p(m)$ and $B$, we aim at deriving in this section the constraints on the mass 
flux $\langle m \rangle$ that an average cloud can transport based on the boundary layer energetics. As will be 
shown in section~6,  $\langle m \rangle$ is the key parameter through which the difference between the mass 
flux distributions of the two reference cases is set. 

We start from the concept of atmospheric convection as a natural heat engine \citep{renno-ingersol:1996}. 
During a heat cycle of an average convective cloud, the heat $Q_{in}$~[J] is input near the surface in the form of 
the turbulent surface heat flux $F_{in}$~[Wm\textsuperscript{-2}] (sum of latent and sensible heat fluxes). This 
heat is partly converted into mechanical work $W_{mech}$ of the convective overturning in the subcloud layer, 
and the rest is added into the cloud layer and redistributed further. Here, we define the heat cycle for the subcloud 
layer that lies between the surface layer over the warm ocean or land surface and the colder cloud layer above. 

The efficiency of the heat cycle is defined as the ratio of mechanical work and the heat input at the surface
\begin{equation}\label{eta}
 \eta = \frac{W_{mech}}{Q_{in}}
\end{equation}
The theoretical maximum efficiency of the heat cycle in the subcloud layer is the Carnot efficiency, which 
can be defined as 
\begin{equation}
\eta_{max} = \frac{T_{sfc}-T_{lcl}}{T_{sfc}}
\end{equation}
$T_{sfc}$ is the surface temperature and $T_{lcl}$ is the temperature at the lifting condensation level. If the heat 
input at the surface would happen solely in form of the sensible heat flux and if no heat was spent to transport 
water vapor out of the subcloud layer, the efficiency of the convective heat cycle would approach the Carnot efficiency. 
However, the thermodynamic cycle of convection in the boundary layer is a mixed moist heat cycle with an efficiency 
that is lower than the maximum theoretical Carnot efficiency, $\eta<\eta_{max}$. As shown in \cite{shutts-gray:1999}, 
the efficiency of the moist heat cycle can be expressed as (see their Eq.~19)
\begin{equation} \label{neweta}
\eta = \frac{B}{1+B} \left( 1 + \frac{\epsilon c_p T_{sfc}}{L_v B} \right)   \frac{g H }{ 2 c_p T_{sfc}}.
\end{equation}
where $c_p$ is the specific heat capacity of the dry air at constant pressure, $L_v$ is the latent heat of vaporization, 
$g$ is the gravitational acceleration, $\epsilon= 1-R_v/R_d=0.608$, $R_v$ is the gas constant for water vapour, 
$R_d$ is the gas constant for dry air, and $H$ is the subcloud layer depth. Eq.~\ref{neweta} is derived under the 
assumption that the effective heat input at the surface, $\eta F_{in}$, is used to maintain convection against mechanical 
dissipation in a convective system in statistical equilibrium. The efficiency of a moist heat cycle $\eta$ could be further 
explained using the entropy budget analysis as in \cite{pauluis-held:2002}. They found that convection acts both as 
a heat cycle and as an atmospheric dehumidifier, and the irreversible entropy production by the two processes are 
in competition. The more the atmosphere acts as a dehumidifier, the less effective it is to generate kinetic energy of 
convective circulations \citep{pauluis-held:2002}. 

From Eq.~\ref{neweta}, it follows that the Bowen ratio highly influences the fraction of the heat input that can be 
transformed into mechanical work to maintain convective circulations. $B$ appears explicitly in Eq.~\ref{neweta} 
but also implicitly through its control on the depth of the subcloud layer $H$ \citep[see][]{schrieber:1996, stevens:2007}. 

Equation~\ref{eta} does not explicitly relate $\langle m \rangle$ to the moist heat cycle. To do so, we proceed as 
follows. The average cloud-base mass flux per cloud $\langle m \rangle$ is related to the turbulent flux of the 
moist static energy at the cloud-base level $\rho \overline{w'h'}$ through the mass flux approximation as defined 
in \cite{arakawa-schubert:1974}:    
\begin{equation}\label{mfappr}
\rho \overline{w'h'} \approx \sum_i m_i (h_{i} - \overline h)  
\end{equation}
where $i=1,...,N$ is the index of individual clouds, and $h_i - \overline h$ is the excess of the moist static energy 
within the updrafts that form clouds with respect to the environment, and an overline denotes averaging over the 
domain.  

As a first simple hypothesis, we assume that the turbulent flux of the moist static energy at cloud base, $\rho \overline{w'h'} $,  
is proportional to the effective surface forcing of the cloud ensemble, $N \eta F_{in}$, and by using Eq.~\ref{mfappr} we write
\begin{equation}\label{wh}
\sum_i m_i (h_{i} - \overline h)  \approx C_1 ~N \eta F_{in}
\end{equation}
where $C_1$ is a proportionality constant, which can be seen as a factor of correction for further heat losses not taken 
into account, and $N$ is the number of clouds in the cloud ensemble. Because the surface forcing is homogeneous, 
$F_{in}$ is equal for all individual cloud heat cycles. The efficiency is controlled by the homogeneous surface properties 
and the subcloud layer depth and is approximately equal among the clouds (see Eq.~\ref{neweta}), so $\eta$ is treated 
as a constant in a single convective case. Now we apply the mass-flux-weighted averaging as defined in \cite{yanai:1973} 
to Eq.~\ref{wh} 
\begin{equation}
\frac{ \sum_i   m_i  h_{i}}{\sum_i   m_i } \approx \tilde h 
\end{equation}
 $\tilde h $ is the mass-flux-weighted average of $h$, which is approximately equal to the average of the moist static 
 energy per cloud, $\langle h \rangle$, where the brackets $\langle . \rangle$ denote averaging over the cloud ensemble.  
 The relative difference between the values of $\tilde h $  and $\langle h \rangle$  is lower than 0.5~\% as estimated from 
 LES. So, we  can rewrite the left-hand side of Eq.~\ref{wh} as
\begin{equation}
\sum_i   m_i  (h_{i} - \overline h) = \sum_i   m_i  h_{i} - \sum_i  m_i \overline h = \frac{\sum_i   m_i  \sum_i   m_i  h_{i}}{\sum_i   m_i } - \sum_i  m_i \overline h \approx  \sum_i   m_i  (\langle h \rangle - \overline h)
\end{equation}
\begin{equation}
\sum_i   m_i   (\langle h \rangle - \overline h)  \approx C_1 ~N \eta F_{in}
\end{equation}
An average moist heat cycle per cloud can then be expressed as 
\begin{equation} \label{avg-cyc}
\langle m \rangle  (\langle h \rangle - \overline h) \approx  C_1 ~\eta F_{in}
\end{equation}
and the average mass flux per cloud is then approximately equal to
\begin{equation} \label{m-cyc}
\langle m \rangle  \approx C_1 ~\frac{\eta F_{in}  }{\langle h \rangle - \overline h    }.
\end{equation}
with $\eta$ given by Eq.~\ref{neweta}.

\begin{figure*}[t]
			{\includegraphics[width=1\linewidth, clip=true, trim= 10mm  50mm 10mm 50mm]{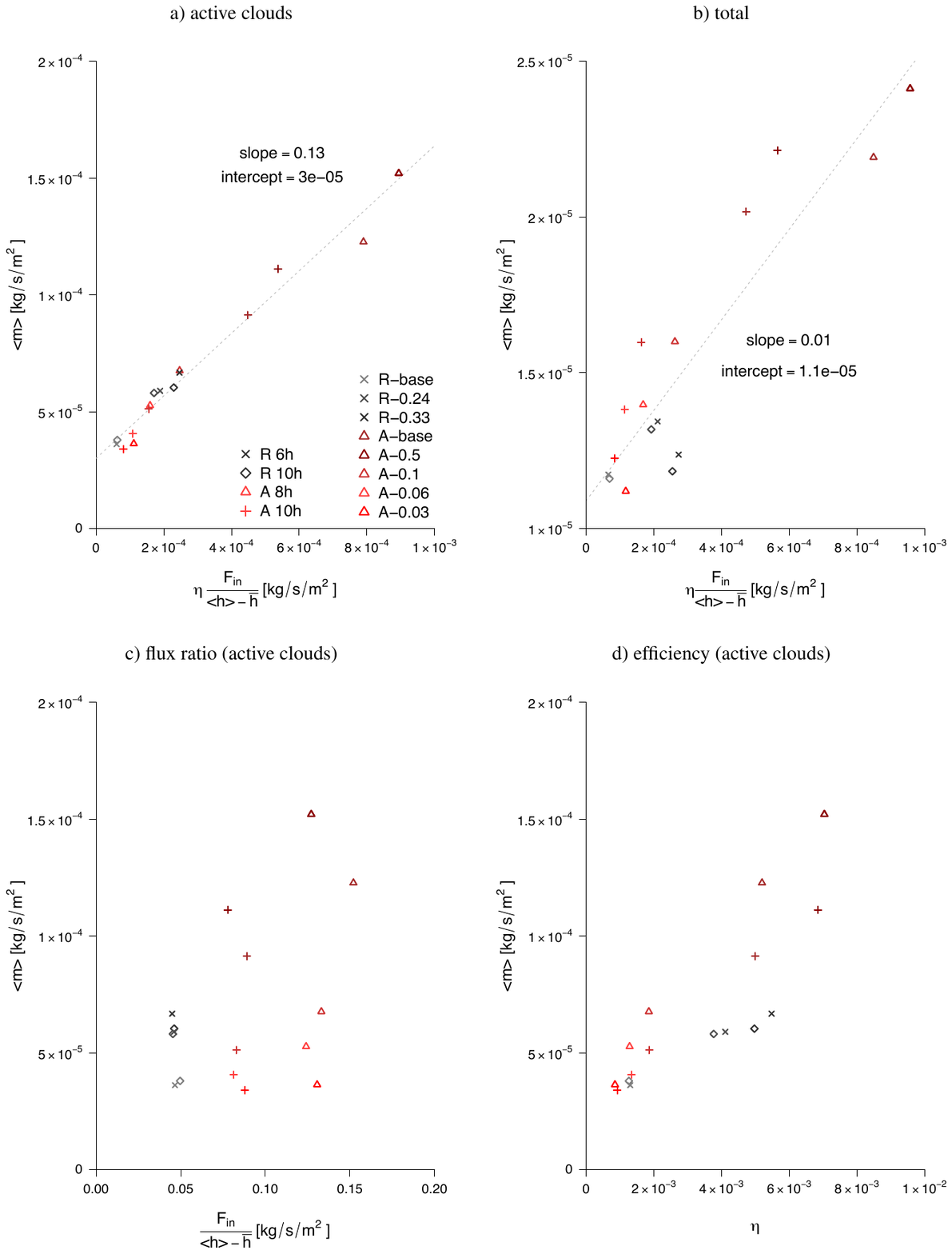}} \\
	\caption{Scaling of the average cloud base mass flux $\langle m \rangle$ based on the moist heat cycle (Eq.~\ref{m-cyc}) 
		for eight LES cases showing 
		a) the average mass flux of active cumulus clouds, b) the average mass flux of the total cloud ensembles, and the right hand 
		side of the Eq.~\ref{m-cyc} decomposed into 
		c) ${ F_{in}  }/({\langle h \rangle - \overline h   })$ and d) $\eta$. Two time frames are used for these figures, the frames 
		starting after 6 and 10 hours of the simulation in the RICO based cases, and after 8 and 10 hours in the ARM based cases. }
	\label{hcyc}
\end{figure*}

We look into the LES simulations to find evidence to support Eq.~\ref{m-cyc}. We base our analysis on the active 
cloud group and we plot the average mass flux per active cloud $\langle m \rangle$ versus the right-hand side 
of the equation~\ref{m-cyc} (Fig.~\ref{hcyc}a). It turns out that Eq.~\ref{m-cyc} holds remarkably well for the 
eight tested LES cases of Table~\ref{cases}, which suggests that the average mass flux per cloud is determined 
by the moist heat cycle of the subcloud layer. The coefficient of determination of a linear regression model is 
$r^2=0.95$. The slope is estimated to be equal to $C_1=0.13$. The intercept parameter is nevertheless not 
equal to zero and results in an additional mass flux which we will denote by $m_{0}$:
\begin{equation}
\langle m \rangle  = m_{0} + C_1 ~ \frac{\eta F_{in}  }{\langle h \rangle  - \overline h  } 
\end{equation}
The estimated value in this study is $m_{0} = 3 \cdot 10^{-5} $~kg/s/m\textsuperscript{2}. 
Depending on the test case and the Bowen ratio value, $\langle m \rangle$ can be 1.5 to 6.9 times larger than 
$m_{0}$ (Fig.~\ref{hcyc}a). 

The scaling Eq.~\ref{m-cyc} is evaluated in Fig.~\ref{hcyc}a only for the active clouds, while we do not show the 
scaling for the "passive" cloud group. This is because the buoyancy threshold used to separate the clouds into 
the two groups misinterpret some active clouds as passive. We can however show the scaling for the total cloud 
ensemble in Fig.~\ref{hcyc}b, which still holds.  

Equation~\ref{m-cyc} is decomposed into two parts to test the dependency of $\langle m \rangle$ of the active 
clouds on $\frac{ ~F_{in}  }{\langle h \rangle - \overline h  } $ and $\eta$ separately (Fig.~\ref{hcyc}c,d). It is clear 
from Fig.~\ref{hcyc}c that $\langle m \rangle$ does not scale with $\frac{ ~F_{in}  }{\langle h \rangle - \overline h } $. 
The points are aligned vertically in three different groups associated with the three main values of the ratio 
$\frac{ ~F_{in}  }{ \langle h \rangle - \overline h  }$, i.e. 0.05, 0.08, and 0.13. The increase in $\langle m \rangle$ 
in each of these three groups of points is due to changing values of $B$. Thus, $\langle m \rangle$ is controlled 
by $B$, while the different mean states of the subcloud layer can still result in the same value of $\langle m \rangle$.
It is not shown here explicitly, but $\langle m \rangle$ also does not scale uniquely with the total surface heat flux $F_{in}$. 

The average mass flux per cloud $\langle m \rangle$ is also not uniquely determined by $\eta$. $\eta$ sets the slope of the three 
lines corresponding to the three different magnitudes of the ratio $\frac{ ~F_{in}  }{\langle h \rangle - \overline h  }$. Furthermore, 
if the ratio $\frac{ ~F_{in}  }{\langle h \rangle - \overline h  }$ in a given group of points is higher, the efficiency $\eta$ in the same 
group is lower compared to the other groups. As a result, $\langle m \rangle$ is uniquely determined by the product of the two factors 
(Eq.~\ref{m-cyc}), with $\eta$ playing the key role in setting the dependence on $B$. 

The fact that $B$ sets the efficiency of the moist convective heat cycle, and thus also controls the expected value of $p(m)$, 
directly explains why the magnitude of the surface forcing does not influence the distribution shape. From this it also follows 
that changes of the surface forcing over the diurnal cycle can not alter the distribution shape, as long as $B$ does not change 
significantly over the diurnal cycle. The moist heat cycle formalism might also explain why self-organization is not a powerful 
driver for the distribution $p(m)$. The convecting system is forced by the same amount of heat input and the efficiency of the 
moist heat cycle is the same at all stages of cloud organization. So, for the shape of $p(m)$, the spatial distribution of clouds 
does not play any significant role. 

An important point to notice here is that the heat cycle formalism applies to the average convective cycle per cloud, and thus 
it determines $\langle m \rangle$, not the bulk contribution $M$ of all clouds in the shallow cloud ensemble. $M$ is not fully constrained 
by the heat cycle of the subcloud layer. This has an implication for the bulk closure assumptions in parameterization of 
convection. To retrieve the closure of a bulk parameterization, an average mass flux per cloud $\langle m \rangle$ 
has to be multiplied by the total number of clouds $N$ to result in the bulk mass flux $M$. Therefore, the controls on $M$ 
may be decomposed into two contributions: the surface conditions control $\langle m \rangle$, while in addition to the surface conditions 
the large-scale forcing acts to set the number of clouds in the ensemble $N$. 
 
\section{Parameters of the mass flux distribution} 

For the application to parameterizations based on the spectral cloud ensembles \citep[][]{arakawa-schubert:1974} 
or stochastic cloud ensembles \citep[as in ][]{pc:2008, sakradzija:2015}, a functional form for $p(m)$ has to be defined 
and the corresponding distribution parameters have to be estimated. In the following, we adopt the mixed Weibull distribution
as a functional form for $p(m)$ as in \cite{sakradzija:2015} 
\begin{equation} \label{weibull}
\begin{split}
p(m) = (1-f) \frac{k_p}{\lambda_p} \left( \frac{m}{\lambda_p} \right)^{k_p-1} e^{-\left( m/\lambda_p \right)^{k_p}} + \\ 
f \frac{k_a}{\lambda_a} \left( \frac{m}{\lambda_a} \right)^{k_a-1} e^{-\left( m/\lambda_a \right)^{k_a} }
\end{split}
\end{equation}
where $f$ is the fraction of active cumulus clouds, $k$ is the shape and $\lambda$ is the scale parameter of the Weibull 
distribution, and subscripts $_p$ and $_a$ denote passive and active distribution modes. 

From the results of the previous section we know that $\langle m \rangle$ varies with the surface conditions. The question 
nevertheless remains whether any of the remaining distribution parameters, namely $k_{p,a}$ and $\lambda_{p,a}$, are universal 
constants. In the study of \cite{sakradzija:2015} these parameters were estimated only for the RICO case for the time period 
of six hours, starting after six hours of simulation. The estimated shape parameter was $k_p=k_a=0.7$ for the given cloud 
sample. In the following, we extend the analysis over longer time period of the RICO case, and over land conditions in the 
ARM case. In the following we focus on the estimation of the shape parameters, $k_{p,a}$, while the scale parameters of 
the Weibull distribution modes, $\lambda_{p,a}$, can be calculated from the expected value of the distribution, 
$\langle m \rangle_{p,a} = \lambda_{p,a} \Gamma[1+1/k_{p,a}]$.  

\begin{figure}[h]
	\centering
		{\includegraphics[width=1\linewidth, clip=true, trim= 0mm  70mm 0mm 70mm]{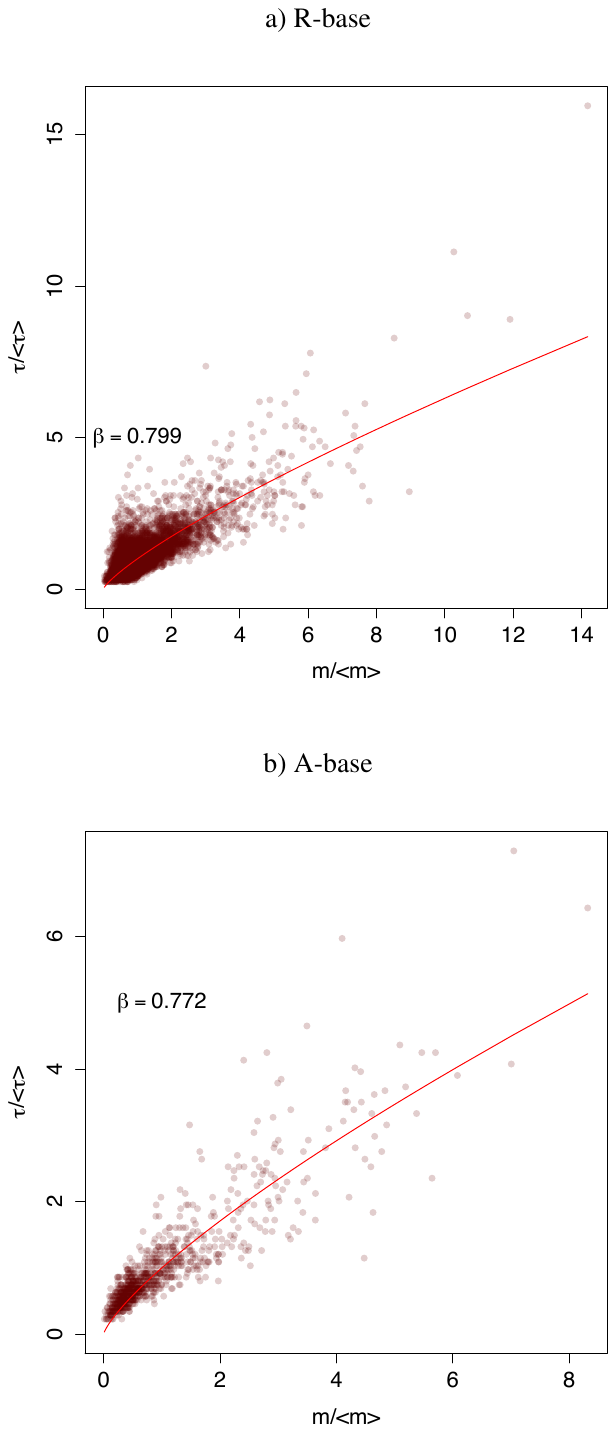}} \\
	\caption{ Scatter plot of individual active cloud lifetimes and cloud base mass fluxes normalized by the ensemble 
		average values: a) R-base simulation and b) A-base simulation. Cloud samples are collected during one hour 
		starting from the 10th~simulation hour. The red line is a fit of the function 
		$\tau / \langle \tau \rangle = (m/\langle m \rangle)^\beta$ using the non-linear least squares. }
	\label{dur-mf}
\end{figure}

In shallow cumulus cloud ensembles, the shape parameter that is less than one, $k<1$, indicates that the memory of 
cloud lifecycles has an effect on the distribution shape \citep{sakradzija:2015}. This effect takes place through 
correlation between the cloud lifetimes $\tau_i$ and the cloud base mass fluxes $m_i$, which is already demonstrated 
for the RICO case in \cite{sakradzija:2015}. We confirm this finding for the RICO case, and we also show that it holds 
in the ARM case (Fig.~\ref{dur-mf}). This correlation is high with the correlation coefficient equal to $r_p = 0.8$ in 
RICO and $r_p = 0.9$ in ARM, estimated for the active cumulus clouds. We assume that this correlation can be described 
by a power law relation $\tau / \langle \tau \rangle = (m/\langle m \rangle)^\beta$, where $\beta$ is the power 
exponent, and $\langle \tau \rangle$ is the average cloud lifetime, as in \cite{sakradzija:2015}. 

In the theory of extreme events, it is known that long-term correlations with a power-law decay of the autocorrelation 
function lead to Weibull distributions of return intervals between rare events \cite[e.g.][]{bunde:2003,blender:2008}. 
In that case the power-law exponent of the autocorrelation function, $t^{-\beta}$, can be assumed equal to the shape 
parameter of the Weibull distribution, $k$  \citep[e.g.][]{blender:2008}. Following this reasoning, the normalized lifetime 
expression $\tau / \langle \tau \rangle = (m/\langle m \rangle)^\beta$ also leads to a Weibull distribution for the 
cloud base mass flux distribution (Eq.~\ref{weibull}). The power-law exponent can then be related to the shape parameter 
of the active mode of the Weibull distribution as $ k_a \approx \beta$. The nonlinear least square fit in Fig.~\ref{dur-mf} 
gives the values for the exponent $\beta=0.8$ in RICO and $\beta=0.77$ in ARM. Hence, it appears that $\beta$ is 
independent on the case set-up. The passive cloud group is more dispersive (not shown here) and the statistical fit is 
thus more uncertain, however we will assume that $k_p \approx k_a = 0.8$. 

\begin{figure*}[p]
	\centering
			{\includegraphics[width=1\linewidth, clip=true, trim= 10mm  90mm 10mm 80mm]{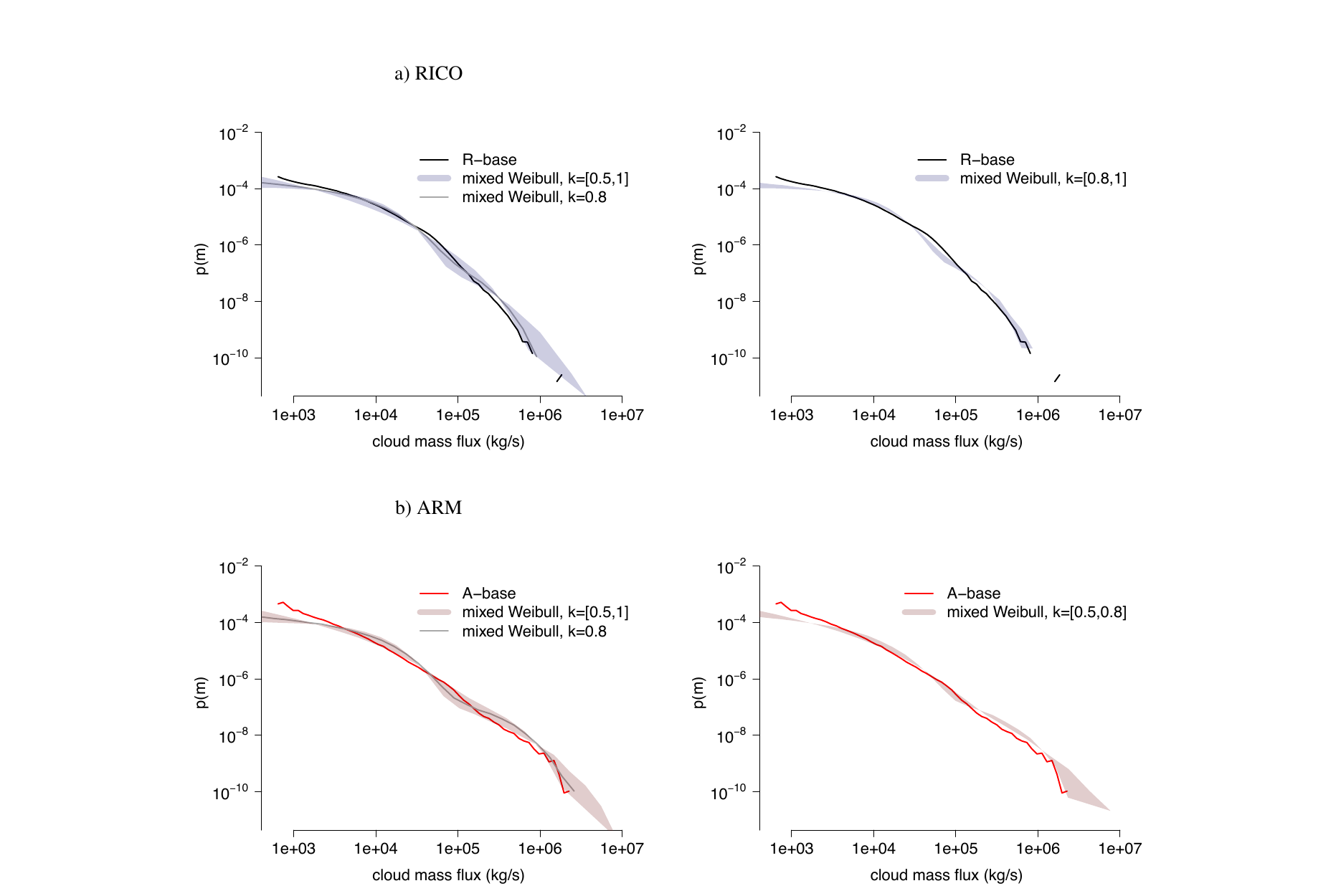}} \\
	\caption{The mass flux distribution approximately fitted using a bimodal Weibull function. 
		The distribution fit of RICO is shown in the upper plot a), while the distribution fit of ARM is shown 
		in the lower plot b). The range of the shape parameter $k$ is quite wide to show low sensitivity of the distribution overall 
		shape to this parameter, while the fraction of active clouds in the ensemble is $f=5$\%. }
	\label{fitting}
\end{figure*}

Combination of the two Weibull modes of the same shape parameter $k_p=k_a=k$, but different $\langle m_p \rangle$ 
and $\langle m_a \rangle$, and hence different $\lambda_{a,p}$, can explain the difference between the two cases 
(Fig.~\ref{fitting}). To construct Fig.~\ref{fitting} and in the purpose of highlighting the uncertainties in $p(m)$ due to 
the chosen value of $k$ only, we here calculate the values of $\langle m_p \rangle$ and $\langle m_a \rangle$ directly 
from the LES output rather than using the formalism of a thermodynamic cycle. The chosen value of $k_p=k_a=0.8$ 
provides a good fit to both distributions (Fig.~\ref{fitting}). On the same plot, we also test a broad range of values for 
$k$, which demonstrates that $k$ is of secondary importance in determining the final shape of $p(m)$. It is evident that $k$ 
can still take a wide range of values, [0.8,1] for RICO and [0.5,0.8] for ARM, for the correct reproduction of the distribution $p(m)$. 
Therefore, we conclude that the main parameter that sets the difference in $p(m)$ among the shallow cumulus cases 
is $\langle m \rangle$. 

The parameter $f$, which is the proportion that active clouds take in the cloud ensemble, is about 4~to~5~\% of 
the total cloud population both in ARM and RICO. This is valid for the distribution of lifetime average mass 
fluxes during time frames of one hour duration, and including only those clouds that are initiated during the time frames.
We choose here to set the value of $f$ to 0.05. 

\section{Conclusions}

The probability distribution of cloud base mass flux $p(m)$ differs among shallow cumulus cases. These 
differences manifest themselves through various shapes, slopes and scales of the distribution. 
Based on the examination of one typical LES case over the ocean (RICO) and one typical LES case over land 
(ARM), and nine variations of these two cases, we propose an explanation for the differences in $p(m)$ 
among shallow cumulus cases. 

The set-up of the two reference LES cases differs in the strength and partitioning of the surface turbulent 
heat fluxes, as well as in the prescribed large-scale forcing tendencies. The ARM case has a strong diurnal 
cycle that is typical for land conditions, while there is no diurnal cycle in the simulation over the ocean (RICO). 
In addition, the cloud field in the RICO case is strongly organized, with manifestation of cold pools and arc 
structures. We have investigated which of these differences in the LES set-up is responsible for the distinct 
shapes of the distribution $p(m)$. 

Analysis demonstrates that partitioning of the surface turbulent fluxes into sensible and latent heating, 
the Bowen ratio $B$, is the only parameter that controls the shape of the distribution $p(m)$. This control 
appears to be governed by the second law of thermodynamics and can be explained by interpreting moist convection 
in the boundary layer as a combination of moisture and heat cycles \citep[as in ][]{shutts-gray:1999, pauluis-held:2002}. 
The efficiency of the moist heat cycle, $\eta$, is less than the Carnot cycle efficiency, because it is directly set by 
the surface Bowen ratio and the depth of the convecting layer \citep{shutts-gray:1999}. Through $\eta$, 
the Bowen ratio controls the average mass flux per cloud $\langle m \rangle$. 

Using the formalism of a moist heat cycle, a scaling law for $\langle m \rangle$ is derived (see Eq.~\ref{m-cyc}). 
By this scaling, the average vertical mass flux through cloud base $\langle m \rangle$  is proportional to the ratio 
of the effective surface heat flux $\eta F_{sfc}$ and the excess in the moist static energy at the cloud base with respect 
to the environment $\langle h \rangle - \bar{h}$. This scaling holds remarkably well for the active buoyant clouds in the eight considered convective 
cases, and thus suggests an universal law across a wide range of the control parameter $B$. Passive and forced clouds 
are not investigated here due to their uncertain separation from the active clouds, but we show that the scaling still holds 
considering all cloud types. 

As such, $B$ controls the shape of the distribution $p(m)$ through its control on $\langle m \rangle$. 
We have demonstrated that different shapes of the distribution $p(m)$ can be well captured by a two-mode 
Weibull distribution function. The shape parameter of each distribution mode is $k<1$ and it is of secondary 
importance for determining the final shape of $p(m)$. The reason for this robustness comes from similarity 
of the power-law exponent $\beta$ in the relation between cloud lifetime and cloud-base mass flux across the LES cases. 
This power-law exponent sets the unique value of the shape parameter across the LES cases.

The Bowen ratios tested in this study covered the range of values between 0.03 and 0.5. This range 
corresponds to the span of conditions covering ocean surfaces to temperate forests and grasslands. In order to 
make the conclusions of this study more general, it would be advantageous to expand the study to dry land surfaces 
and to extend the analysis to cloud observations. In addition, the mechanical forcing in the two reference cases 
was of similar magnitude. A question left for further investigation is how stronger winds and higher wind shears 
might influence the convective mass flux and population statistics. 

One of the key outcomes of this study is that the concept of a moist heat cycle applies to an average convective 
cloud cycle. In order to retrieve the total mass flux in a cloud ensemble $M$, it is necessary to set the constraints 
on the number of clouds $N$ in every given case, since $M=N\langle m \rangle$. $N$ does not appear to be 
constrained by the moist heat cycle. One may hypothesize that $M$ is governed by the large-scale forcing through 
control on the number of clouds $N$, in addition to the surface conditions that impose a constraint on $\langle m \rangle$.  

The results of this study also have implications for the cloud size distribution, which has a very similar shape to 
the distribution $p(m)$.  Various shapes and slopes of the cloud size distribution that are observed and have been 
documented in literature, may just reflect changes in Bowen ratios encountered across various studies. The various 
proposed distribution shapes could be encompassed by a single functional form given by the mixed Weibull distribution 
function. Such a multimodal distribution function already encompasses all the observed shapes, starting from an 
exponential shape to power-laws, depending on the value of the distribution parameters. Based on this study, the 
expected value of the cloud size distribution might impose the only relevant control on the distribution shape, which 
then could be constrained by the underlying physical processes in the boundary layer.

%
\acknowledgments
This research was carried out as part of the Hans-Ertel Centre for Weather Research. 
This research network of Universities, Research Institutes and the Deutscher Wetterdienst 
is funded by the BMVI (Federal Ministry of Transport and Digital Infrastructure). 
Primary data and scripts used in the analysis and other supplementary information 
that may be useful in reproducing the author's work are archived by the Max Planck 
Institute for Meteorology and can be obtained by contacting publications@mpimet.mpg.de. 
Computation is performed using the Detutsches Klimarechenzentrum (DKRZ) resources. 
Plots are made using the R statistical software. 
The authors thank Juan-Pedro Mellado and Alberto de Lozar for fruitful discussions and comments 
on an early version of this manuscript. We are also grateful to two anonymous reviewers for their constructive 
comments and suggestions that helped improving the manuscript. 

%



%
\begin{table*}[h]
	\caption{List of the LES cases with the abbreviations used in the text, the case on which the simulations are based on, 
		the maximum Bowen ratio $B_{max}$, the total surface turbulent heat flux averaged over the simulation period 
		$\overline F_{in} = \overline {F_{sh}+F_{lh}} $, 
		and the type and duration of the large-scale forcing $LS_{forc}$. }\label{cases}
	\centering
	\begin{tabular}{lcccccc}
		\hline
		abbr. & reference case  & $B_{max}$                      & $\overline F_{in} [W/m^2]$   & $LS_{forc} $          \\
		\hline
		R-base           & RICO          & 0.06                          &  171                                          &    $const.$          \\
		R-0.24           & RICO          & 0.24                          &  152                                          &   $const.$          \\
		R-0.33           & RICO          & 0.33                          &  152                                          &   $const.$          \\
		A-base           & ARM          & 0.36                          &  343                                          &       14 h 30'       \\
		A-0.5             & ARM          & 0.50                          &  340                                          &       14 h 30'       \\
		A-0.1             & ARM          & 0.11                          &  347                                         &       14 h 30'       \\
		A-0.06           & ARM          & 0.06                          &  348                                          &       14 h  30'      \\
		A-0.03           & ARM          & 0.03                          &  349                                          &       14 h   30'     \\
		A-lowflx        & ARM          & 0.36                          &   274                                         &       14 h    30'    \\
		A-short          & ARM          & 0.36                          &  341                                          &       10 h            \\        
		A-long           & ARM          & 0.36                          &  344                                          &       19 h            \\
		\hline 
	\end{tabular}
\end{table*} 

%

\end{document}